\documentclass[reprint,aps,twocolumn,prmaterials,superscriptaddress]{revtex4-2}

\usepackage{gensymb}
\usepackage{amssymb}
\usepackage{bm, amsmath}
\usepackage{bm}
\usepackage{physics}

\usepackage[euler]{textgreek}
\usepackage{siunitx}
\usepackage[colorlinks=true,linkcolor=blue,urlcolor=blue,citecolor=blue]{hyperref}
\usepackage[version=4]{mhchem}
\usepackage{natbib}
\usepackage{siunitx}

\usepackage{graphicx}
\usepackage[caption=false]{subfig}
\usepackage[outdir=./]{epstopdf}

\usepackage{booktabs}
\usepackage{dcolumn}
\usepackage{multirow}

\usepackage{orcidlink}

\usepackage[normalem]{ulem} 
\usepackage{verbatim}
\usepackage[draft]{todonotes}
\usepackage{float}
\usepackage{color}

\newcommand{\SIdash}[2]{\SI[number-unit-product={\text{-}}]{#1}{#2}}
\DeclareSIUnit{\QL}{QL}
\DeclareSIUnit{\Torr}{Torr}
\DeclareSIUnit{\bohr}{\mu_{\mathrm{B}}}
\DeclareSIUnit{\fu}{f.u.}
\begin{document}
\title{Structure and magnetism of MnGe thin films \\grown with a nonmagnetic CrSi template}

\author{B.~D.~MacNeil\,\orcidlink{0000-0002-4731-3528}}
\affiliation{Department of Physics and Atmospheric Science, \href{https://ror.org/01e6qks80}{Dalhousie University}, Halifax, Nova Scotia B3H 3J5, Canada}

\author{J.~S.~R.~McCoombs\,\orcidlink{0009-0005-6955-6335}}
\affiliation{Department of Physics and Atmospheric Science, \href{https://ror.org/01e6qks80}{Dalhousie University}, Halifax, Nova Scotia B3H 3J5, Canada}

\author{D.~Kalliecharan\,\orcidlink{0000-0002-6429-279X}}
\affiliation{Department of Physics and Atmospheric Science, \href{https://ror.org/01e6qks80}{Dalhousie University}, Halifax, Nova Scotia B3H 3J5, Canada}

\author{J.~Myra\,\orcidlink{0009-0002-3087-2013}}
\affiliation{Department of Physics and Atmospheric Science, \href{https://ror.org/01e6qks80}{Dalhousie University}, Halifax, Nova Scotia B3H 3J5, Canada}

\author{M.~Pula\,\orcidlink{0000-0002-4567-5402}}
\affiliation{Department of Physics and Astronomy, \href{https://ror.org/02fa3aq29}{McMaster University}, Hamilton, Ontario L8P 4N3, Canada}

\author{J.~F.~Britten\,\orcidlink{0000-0001-5527-0620}}
\affiliation{McMaster Analytical X-ray Diffraction Facility, \href{https://ror.org/02fa3aq29}{McMaster University}, Hamilton, Ontario L8P 4N3, Canada}

\author{G.~B.~G.~Stenning}
\affiliation{\href{https://ror.org/01t8fg661}{ISIS Neutron and Muon Source}, STFC Rutherford Appleton Laboratory, Didcot OX11 0QX, United Kingdom}

\author{K.~Gupta\,\orcidlink{0000-0002-4826-6620}}
\affiliation{\href{https://ror.org/00k1qja49}{Catalan Institute of Nanoscience and Nanotechnology (ICN2)}, Campus UAB, Bellaterra, Barcelona 08193, Spain}

\author{G.~M.~Luke\,\orcidlink{0000-0003-4762-1173}}
\affiliation{Department of Physics and Astronomy, \href{https://ror.org/02fa3aq29}{McMaster University}, Hamilton, Ontario L8P 4N3, Canada}

\author{T.~L.~Monchesky\,\orcidlink{0000-0001-7401-7866}}\thanks{Contact author: tmonches@dal.ca}
\affiliation{Department of Physics and Atmospheric Science, \href{https://ror.org/01e6qks80}{Dalhousie University}, Halifax, Nova Scotia B3H 3J5, Canada}
\date{March 3, 2026}

\begin{abstract}
	{We report a method to grow B20 MnGe thin films using molecular beam epitaxy which employs an ultrathin CrSi template layer on Si(111). This layer is expected to be nonmagnetic, in contrast to MnSi and FeGe buffer layers that have been used previously. This template layer permits an investigation of the intrinsic properties of MnGe in the ultrathin film limit without the influence of a neighboring magnetic layer. Single-phase MnGe(111) films were grown with thicknesses between 2 and \SI{40}{\nano\metre}, which exhibited low interfacial roughnesses on the order of \SI{0.6}{\nano\metre}. The films crystallized in a B20 structure with a small rhombohedral distortion. Magnetometry measurements in out-of-plane fields are consistent with a cone phase derived from helimagnetic order propagating along the film normal. However, an unexpected remanent moment develops below \SI{35}{\kelvin}, concomitant with features in the field dependence of the transport data. This provides indirect evidence for the presence of a low-temperature phase which has been identified by others as either a triple-$Q$ topological spin-hedgehog lattice, or a multi-domain single-$Q$ helical state.}
\end{abstract}

\maketitle

\section{Introduction}
The B20 monosilicides and monogermanides host a rich phase diagram of magnetic states due to their broken inversion symmetry. The competition between symmetric exchange and the Dzyaloshinskii-Moriya interaction (DMI) results in helical magnetic order, with additional regions of the phase diagrams that host exotic magnetic textures~\cite{Dzyaloshinskii_1958, Moriya_1960, BakJensen_1980, Bogdanov_1989}. The prototypical B20 magnets MnSi and FeGe are known to host a skyrmion state near their ordering temperature, where the magnetization twists into nanoscale cylindrical structures that are topologically protected~\cite{Muhlbauer_2009, Yu_2011}. The relative strength of symmetric exchange and DMI sets the wavelength of the helix and the lengthscale of the magnetic textures, which are typically much larger than the atomic spacing, e.g., approximately \SI{18}{\nano\metre} for MnSi and \SI{70}{\nano\metre} for FeGe.

On the other hand, MnGe exhibits the shortest helical wavelength among B20 compounds, which varies between approximately \SI{3}{\nano\metre} at low temperature, up to \SI{6}{\nano\metre} near the ordering temperature~\cite{Kanazawa_2011, Makarova_2012, Kanazawa_2012}. Electronic structure calculations show that the short helical wavelength is not due to DMI alone, but rather further-range exchange interaction~\cite{Mendive_2021}. While skyrmions have not been observed in MnGe, there are reports of a unique topological magnetic phase consisting of three-dimensional magnetic textures called spin hedgehogs~\cite{Tanigaki_2015, Kanazawa_2016, Kanazawa_2017, Kanazawa_2020, Fujishiro_2021, Kitaori_2021, Hayashi_2021, Pomjakushin_2023}. Theoretical models describe it in terms of superposition of three orthogonal helical spin-density waves~\cite{Pomjakushin_2023}, referred to as a triple-$Q$ state, that can arise due to biquadratic exchange coupling \cite{Mendive_2021} or a novel topological chiral interaction~\cite{Grytsiuk_2020}. The presence of this state was inferred from a large anomaly in the Hall resistivity attributed to a topological Hall effect, and Lorentz transmission electron microscopy (LTEM) measurements~\cite{Tanigaki_2015}. However, some studies claim that the triple-$Q$ phase is actually a topologically trivial multi-domain single-$Q$ helical state with wavevectors aligned along each of the three $\langle 100 \rangle$ directions~\cite{Makarova_2012, Deutsch_2014, Altynbaev_2016, Yaouanc_2017, Altynbaev_2018, Martin_2019, Repicky_2021}. A multidomain helical phase could also be responsible for the LTEM contrast observed in Ref.~\cite{Tanigaki_2015}, but seemingly does not account for the large anomalies in the transport measurements.

The B20 phase of MnGe is metastable and preparing suitable samples is difficult. Polycrystalline MnGe was first synthesized at high pressure~\cite{Takizawa_1988, Kanazawa_2011} and the largest single crystals synthesized to date were around \SI{50}{\micro\metre} in size~\cite{Kanazawa_2020}. Epitaxial strain has also been shown to be sufficient to stabilize MnGe. The first example utilized a \SIdash{1}{\nano\metre} seed layer of MnSi to grow MnGe on Si(111) using molecular beam epitaxy (MBE)~\cite{Engelke_2013}. Other groups have used a \SIdash{2}{\nano\metre} MnSi template~\cite{Kanazawa_2017, Kanazawa_2020, Fujishiro_2021, Hayashi_2021, Li_2025} or buffer layers of FeGe with thicknesses between 2 and \SI{10}{\nano\metre}~\cite{Ahmed_2017, Repicky_2021}. To date, all MnGe thin films have been grown using either MnSi or FeGe templates, which themselves are magnetic and can therefore influence the magnetic response of the MnGe. Previous studies have mainly focused on relatively thick MnGe films (\SI{80}{\nano\metre} to \SI{3}{\micro\metre}) and while Ref.~\cite{Engelke_2013} studied films with thicknesses below \SI{13.5}{\nano\metre}, the magnetometry data were dominated by the moment of the MnSi template at low temperature.

\begin{figure*}[!htb]
	\centering
	\includegraphics[scale=1]{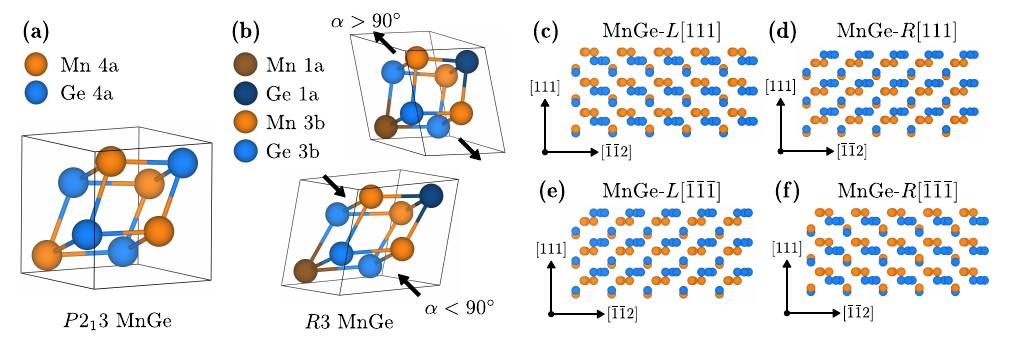}
	\caption{(a) Unit cell of B20 MnGe. (b) Unit cell of rhombohedrally distorted $R3$ MnGe shown for both in-plane tension ($\alpha>\ang{90}$) and compression ($\alpha<\ang{90}$). The arrows indicate the direction of the in-plane strain. (c)--(f) Illustration of the \si{\QL} stacking of the MnGe structure, for the four possible morphologies. The structures (c) and (e) form a left-handed crystal whereas (d) and (f) are right handed. The axes indicate the direction of the local MnGe $[111]$ and $[\bar{1}\bar{1}2]$ directions for each morphology. Crystal structures were visualized using the \textsc{vesta} software~\cite{Vesta}.}
	\label{fig:MnGe_structure}
\end{figure*}

The magnetic structure of MnGe/MnSi/Si(111) films was probed by small-angle neutron scattering (SANS) in Ref.~\cite{Kanazawa_2017}, which revealed single-$Q$ helimagnetic order above \SI{100}{\kelvin}. In contrast to bulk, where the helices aligned along [100], the helical wavevector was along the [111] film normal. Below \SI{100}{\kelvin}, the magnetic diffraction spot along $[111]$ split into two sets of three spots. For the \SIdash{1.8}{\micro\metre}-thick film {($\approx$bulk)}, these spots were along $\langle 100 \rangle$ directions. However, for the 160- and \SIdash{735}{\nano\metre} thick samples, the diffraction features were up to $\sim\ang{37}$ away from $\langle 100 \rangle$ toward the $[111]$. In these thinner samples, the alleged triple-$Q$ state was suppressed to a region below approximately \SI{40}{\kelvin} with the addition of a uniaxial anisotropy due to epitaxial strain from the substrate. Another report using spin-polarized scanning tunneling microscopy on an \SIdash{80}{\nano\metre} MnGe/FeGe/Si(111) film found multidomain single-$Q$ states, and no evidence for a triple-$Q$ hedgehog state~\cite{Repicky_2021}. The intersections of these domains create topological objects, but they occur at a density lower than required to explain the features in the Hall-effect data for MnGe.

In this work, we present the development of a B20 bilayer system designed to investigate the intrinsic magnetic properties of MnGe in the ultrathin film limit, where the helical wavelength is comparable to the film thickness. This system enables exploration of the interplay between finite size and large spin fluctuations, as well as the unique electronic structure of MnGe in relation to its deviation from canonical B20 helimagnetism. We have characterized the structure of MnGe(111) grown on B20 CrSi template layers deposited on Si(111) substrates, and investigated their magnetic properties with magnetometry and magnetotransport studies. We chose CrSi based on several studies that find no evidence of permanent magnetic order down to low temperature~\cite{Radovskii_1965, Mishra_2019}. The work in Ref.~\cite{Kousaka_2014} reported a weak moment of less than $0.04\mu_{\mathrm{B}}$ per Cr atom below \SI{20}{\kelvin} and Ref.~\cite{Banik_2020} measured a moment of $0.01\mu_{\mathrm{B}}$ per Cr atom below \SI{2}{\kelvin} in an applied field of \SI{7}{\tesla}. Our magnetometry measurements agree with these findings and indicate that the CrSi films are paramagnetic and therefore not expected to have a large influence on the MnGe layer beyond potential changes to the electronic structure of the interfacial layer.

The rest of this paper is structured as follows: We first review the structure of B20 films on Si(111) substrates, and then describe the methods that we used to produce both a template of B20 CrSi, and B20 MnGe films using this template. We also discuss the effect of temperature and Mn composition on the stability of the B20 MnGe phase. We then outline the structural experiments we conducted to prove that we indeed grew epitaxial MnGe films, and to characterize the strain and distortion in these films due to the Si substrate. Finally, we investigate the magnetic properties of our films by presenting complementary magnetometry and transport data with a field applied parallel to MnGe$[111]$.

\section{Crystal Structure}
The crystal structure of B20 MnGe (space group No. 198 $P2_13$) is detailed in Figure~\ref{fig:MnGe_structure}. The Mn and Ge atoms are located at the $4a$ Wyckoff sites and form a helix which winds around the $[111]$ directions according to the chirality of the crystal. The stacking of atoms in this direction can be described in terms of a quadruple layer (\si{\QL}), which has a thickness corresponding to the (111) plane spacing of \SI{0.277}{\nano\metre}. A \si{\QL} consists of two dense and two sparse alternating layers of Mn and Ge atoms. The stacking unit along $[111]$ is composed of 12 atomic layers, which can be described as $ABC$ stacking of the \si{\QL}s. The stacking manner is determined by both the crystal chirality and its orientation. The MnSi/Si(111) system was found to nucleate domains with differing stacking sense, which are related by either inversion, or \ang{180} rotation to give $[111]$ or $[\bar{1}\bar{1}\bar{1}]$ pairs~\cite{Trabel_2017, Morikawa_2020}. This yields four possible morphologies for MnGe/Si(111) when accounting for two crystal chiralities and two film orientations, which are denoted as \mbox{MnGe-$L[111]$}, \mbox{MnGe-$L[\bar{1}\bar{1}\bar{1}]$}, \mbox{MnGe-$R[111]$}, and \mbox{MnGe-$R[\bar{1}\bar{1}\bar{1}]$}. These stacking sequences are visualized in Fig.~\hyperref[fig:MnGe_structure]{1(c)-1(f)}.

When a (111)-oriented B20 structure is subject to an isotropic in-plane strain, the cubic structure is distorted to a rhombohedral (space group No. 143 $R3$) symmetry. In the distorted unit cell, the $4a$ Wyckoff sites of the Mn and Ge split into two inequivalent $1a$ and $3b$ Wyckoff sites. The rhombohedral distorted unit cell is depicted in Fig.~\hyperref[fig:MnGe_structure]{1(b)}. Such a distortion was measured in thin film MnSi and \ce{Fe_{1-x}Co_xSi} grown on Si(111) with rhombohedral angles $\alpha$ of \ang{90.5} and \ang{92.5}, respectively~\cite{Karhu_2012, Porter_2012}. The distortion for FeGe was $\alpha=\ang{89.7}$ when grown on Si(111)~\cite{Turgut_2017} but $\alpha=\ang{94}$~\cite{Budhathoki_2020} when grown on Ge(111). The in-plane strain for these systems is expected to be tensile ($\alpha>\ang{90}$), whereas the expected lattice mismatch for MnGe(111)/Si(111) would be \mbox{$(2d_{1\bar{2}1}^{\mathrm{\,film}} - d_{1\bar{1}0}^{\,\mathrm{sub}})/2d_{1\bar{2}1}^{\mathrm{\,film}}=+1.9\%$}, meaning pseudomorphic MnGe would be under compression.

\section{Film Growth}
MnGe thin films were prepared on high resistivity \mbox{($\rho>\SI{100}{\ohm\metre}$)} Si(111) substrates. The wafers were ultrasonically degreased for 15 min each in acetone and methanol, followed by a rinse and overflow in deionized nanopure water. The native oxide was etched for 10 min in a 1:2:10 solution of \ce{NH4OH}, \ce{H2O2} and nanopure water at \SI{75}{\degreeCelsius}, leaving the Si surface terminated by \ce{SiO2}. The wafer was again rinsed and overflowed, then dried with nitrogen gas and immediately loaded into the MBE system with base pressure less than \SI{5e-9}{\Pa}. The deposition rates from the Si, Ge and metal sources were monitored by three independent quartz-crystal microbalances.

\begin{figure}[!htb]
	\centering
	\includegraphics[scale=1]{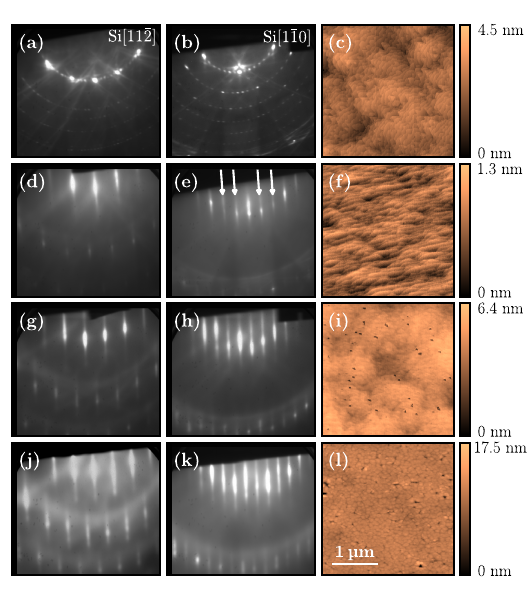}
	\caption{\textit{In situ} RHEED and \textit{ex-situ} AFM images of (a)--(c) Si(111) substrate, (d)--(f) \SIdash{2}{\QL} \ce{CrSi} template (\SIdash{20}{\nano\metre} Si overlayer), (g)--(i) \SIdash{6.0}{\nano\metre} \ce{CrSi} film (\SIdash{20}{\nano\metre} Si overlayer), and (j)--(l) an \SIdash{7.2}{\nano\metre} \ce{MnGe} film (no Si overlayer) atop a \SIdash{2}{\QL} \ce{CrSi} template. The arrows in (e) indicate the fractional-order diffraction streaks visible in the $\surd 3 \times \surd 3$ pseudoreconstruction of \SIdash{2}{\QL} CrSi with the beam along Si$[11\bar{2}]$. These are not present for the thicker CrSi film in (h). The RHEED patterns in each column share the same orientation relative to the substrate direction annotated in (a) and (b). All micrographs share the same scalebar.}
	\label{fig:rheed_afm}
\end{figure}

The wafer and holder were degassed overnight at \SI{570}{\degreeCelsius} and the \ce{SiO2} was thermally desorbed at \SI{830}{\degreeCelsius} for 60 min. A \SIdash{20}{\nano\metre} Si buffer layer was deposited at \SI{640}{\degreeCelsius}, after which the sample was cooled to room temperature at a rate of less than \SI{1}{\degreeCelsius\per\second}. \emph{In situ} reflection high energy electron diffraction (RHEED) revealed a strong $7 \times 7$ reconstruction in Fig.~\hyperref[fig:rheed_afm]{2(a)} and \hyperref[fig:rheed_afm]{2(b)}, indicative of a pristine Si(111) surface which was also evidenced by \emph{ex situ} atomic force microscopy (AFM) of the surface. The AFM image of the substrate in Fig.~\hyperref[fig:rheed_afm]{2(c)} shows step heights of approximately \SI{0.3}{\nano\metre}, which correspond to the atomic spacing between Si(111) planes.

We adapted the method of Higashi \emph{et al.}~\cite{Higashi_2009} to stabilize a \SIdash{2}{\QL} template of B20 CrSi. A trilayer with nominal thicknesses {$\SIdash{0.5}{\QL}$ Si/$\SIdash{1}{\QL}$ Cr/$\SIdash{0.5}{\QL}$ Si} was deposited at room temperature and then annealed under RHEED observation until the $\surd 3 \times \surd 3$ pseudoreconstruction in Fig.~\hyperref[fig:rheed_afm]{2(d)} and \hyperref[fig:rheed_afm]{2(e)} was achieved, typically after 60 min. The micrograph in Fig.~\hyperref[fig:rheed_afm]{2(f)} reveals a smooth surface with roughness commensurate to that of the substrate.

After the room-temperature deposition of the Higashi trilayer, the Si $7\times 7$ reconstruction was replaced by diffuse RHEED spots representing the uncrystallized template. Upon annealing, sharp fractional-order spots [Fig.~\hyperref[fig:rheed_afm]{2(e)}] appeared when the beam was incident along Si$[1\bar{1}0]$, signaling the formation of B20 CrSi. Typical annealing temperatures were between 330 and \SI{410}{\degreeCelsius} for 30 to 75 min. When the CrSi trilayer was annealed above \SI{420}{\degreeCelsius}, the quality of the RHEED pattern dropped considerably and transmission spots indicated a rougher surface. This behavior is indicative of the formation of C40 \ce{CrSi2} (space group No. 180 $P6_222$) which is reported to form in this temperature range with poor epitaxy until annealed above \SI{1100}{\degreeCelsius}~\cite{Wetzel_1987, Wetzel_1988, Mahan_1993}. Attempts at stabilizing a B20 CrSi template using a solid phase reaction between amorphous Cr and the Si $7\times 7$ surface either resulted in the formation of \ce{CrSi2} as evidenced by RHEED and later corroborated by \emph{ex situ} x-ray diffraction (XRD), or no crystallization seen in RHEED (see the Supplemental Material for additional RHEED images identifying \ce{CrSi2} formation~\cite{supp}). Unless otherwise noted, all films were capped with an amorphous Si or Ge overlayer grown at room temperature, ranging from 4 to \SI{20}{\nano\metre} in thickness.

\begin{figure}[!htb]
	\centering
	\includegraphics[scale=1]{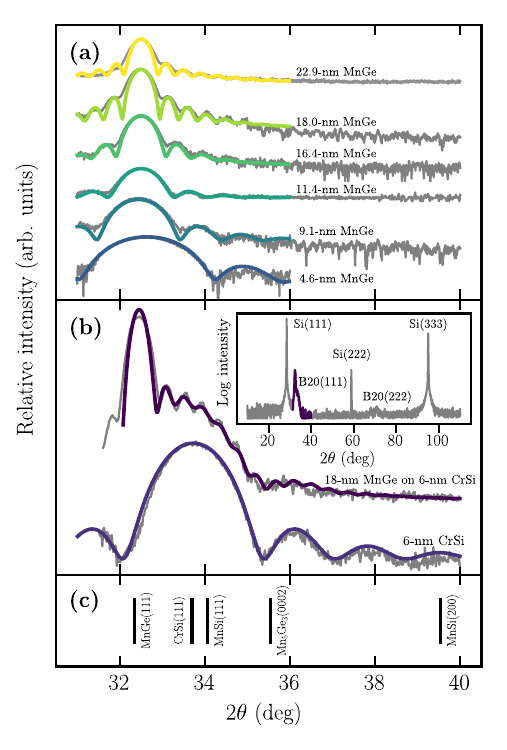}
	\caption{$\theta-2\theta$ XRD measurements of (a) the MnGe (111) peak for films grown atop \SIdash{2}{\QL} CrSi using the two-stage annealing procedure, and (b) the MnGe and CrSi (111) peak for a film grown atop a \SIdash{6}{\nano\metre} CrSi template using the same procedure. The inset in (b) shows phase purity of the B20 structure. Ticks in (c) denote the bulk peak positions for several crystal structures. Fits are performed using the model in Eq.~\ref{eq:xrd_peak}.}
	\label{fig:xrd_MnGe}
\end{figure}

A two-stage annealing recipe was used to produce thicker films of B20 CrSi. Stoichiometric Cr and Si were codeposited onto the \SIdash{2}{\QL} template below \SI{100}{\degreeCelsius} and then annealed to \SI{350}{\degreeCelsius} for approximately 60 min. The sharp diffraction patterns in Fig.~\hyperref[fig:rheed_afm]{2(g)} and \hyperref[fig:rheed_afm]{2(h)} combined with the micrograph in Fig.~\hyperref[fig:rheed_afm]{2(i)} are consistent with a flat surface with roughness below \SI{0.7}{\nano\metre} for a \SIdash{6}{\nano\metre}-thick ($\approx\SI{22}{\QL}$) CrSi film. The XRD peak in Fig.~\hyperref[fig:xrd_MnGe]{3(b)} shows well-resolved Laue fringes with roughness in agreement with AFM. Annealing temperatures exceeding this range would result in the formation of \ce{CrSi2} despite the \SIdash{2}{\QL} template being stable until approximately \SI{410}{\degreeCelsius}.

We found that codeposition of Mn and Ge at lower temperatures ultimately led to smoother films. MnGe films were grown by codeposition of Mn and Ge in a stoichiometric ratio onto the \SIdash{2}{\QL} CrSi template layer at \SI{100}{\degreeCelsius}. At this point, the film produced an amorphous RHEED pattern. Annealing to \SI{250}{\degreeCelsius} for 60--90 min resulted in smooth crystalline surfaces as revealed by the diffraction patterns in Fig.~\hyperref[fig:rheed_afm]{2(j)}-\hyperref[fig:rheed_afm]{2(k)}. A similar method has also been employed with a seed layer of MnSi~\cite{Kanazawa_2017, Kanazawa_2020, Fujishiro_2021, Hayashi_2021, Li_2025}. The sharp streaks and low background are consistent with a flat multidomain surface depicted by the micrograph in Fig.~\hyperref[fig:rheed_afm]{2(l)}. Figure~\hyperref[fig:xrd_MnGe]{3(a)} shows XRD acquired in a Siemens D500 diffractometer using Cu K$\alpha$ radiation for a variety of MnGe films grown with this two-stage annealing procedure. No impurity phases are detectable: only the MnGe (111) and (333) peaks are observed. The \SIdash{6}{\nano\metre} CrSi film also served as a suitable template for MnGe with the same two-stage growth as the \SIdash{2}{\QL} template. A sample XRD measurement of an \SIdash{18}{\nano\metre} MnGe film stabilized on \SIdash{6}{\nano\metre} CrSi is depicted in Fig.~\hyperref[fig:xrd_MnGe]{3(b)}. The inset displays the full XRD pattern of this sample, which shows no evidence of any impurity phases such as \ce{CrSi2}.

We adapted a kinematical model of the Laue oscillations from Ref.~\cite{Boulle_2006} to determine the film thickness and estimate the roughness of the interfaces. The scattering intensity is:
\begin{equation}
	I(q) = \frac{|F|^2}{q^2}\qty[2-2\cos(qd)\exp\qty(-\frac{q^2({\sigma_1}^2 + {\sigma_2}^2)}{2})],
	\label{eq:xrd_peak}
\end{equation}
where $q$ is the scattering vector, $F(q)$ is the form factor, $d$ is the film thickness, and $\sigma_1$ and $\sigma_2$ are the roughnesses at the upper and lower film interfaces, respectively. Accompanying fits to the MnGe(111) Laue fringes in Fig.~\hyperref[fig:xrd_MnGe]{3(a)} yielded interfacial roughnesses below \SI{0.8}{\nano\metre}. We also extracted the roughnesses from X-ray reflectometry (XRR) measurements that were fitted using \textsc{genx}~\cite{GenX} and both methods were in agreement (see the Supplemental Material for XRR data and fitting~\cite{supp}). [See Fig.~\hyperref[fig:params_vs_d]{8(d)}, where we present interfacial roughness of the MnGe layer (film surface) as measured by XRR (AFM)].

\begin{figure}[!htb]
	\centering
	\includegraphics[scale=1]{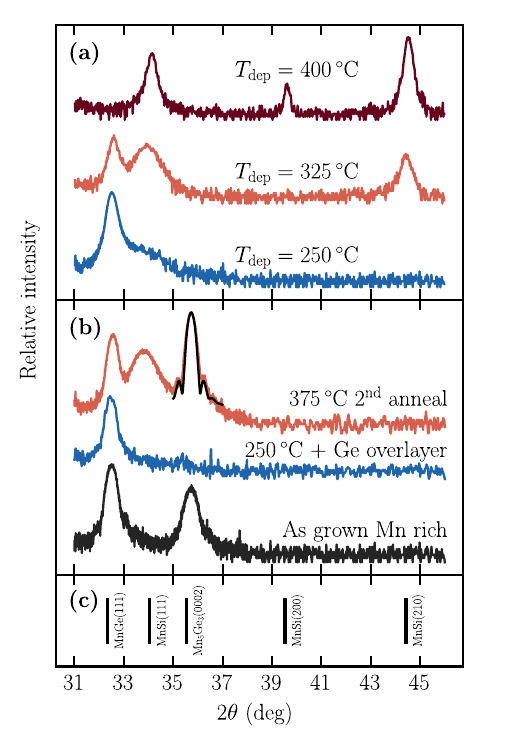}
	\caption{(a) $\theta-2\theta$ XRD measurements comparing the effect of temperature on the deposition of Mn and Ge atop \SIdash{2}{\QL} CrSi, leading to the formation of MnSi by diffusion above \SI{250}{\degreeCelsius}. (b) $\theta-2\theta$ XRD scans for Mn-rich films deposited on \SIdash{2}{\QL} CrSi. The as-grown film exhibits \ce{Mn5Ge3} impurities, which are removed when annealed to \SI{250}{\degreeCelsius} after annealing in excess Ge. This impurity phase remains upon heating above \SI{250}{\degreeCelsius}, where diffusion-formed MnSi is still observed in coexistence with MnGe and (smoother) \ce{Mn5Ge3}. The fit in (b) was performed using the model in Eq.~\ref{eq:xrd_peak}. Ticks in (c) denote the bulk peak positions for several crystal structures.}
	\label{fig:xrd_Mn5Ge3}
\end{figure}

We also attempted to grow MnGe films atop the \SI{2}{\QL} CrSi template using the same MBE recipe employed for MnSi, and the first synthesis of MnGe on Si(111)~\cite{Karhu_2011, Engelke_2013}. When Mn and Ge were co-deposited at \SI{250}{\degreeCelsius} and annealed for 60 min, XRD revealed a single peak at $2\theta = \ang{32.5}$ corresponding to MnGe(111) planes, although the absence of Laue oscillations flanking the main peak indicates that the MnGe film is relatively rough. Furthermore, a broad shoulder is visible near \ang{34} which indicates that some deposited Mn has reacted with the substrate to form MnSi. The same behavior was observed in Ref.~\cite{Engelke_2013} for a \SIdash{1}{\nano\metre}-thick ($\approx\,$\SI{3.8}{\QL}) MnSi template. Increasing the temperature to \SI{325}{\degreeCelsius} led to more diffusion and a coexistence of epitaxial MnGe and textured MnSi phases which is replaced solely by the textured MnSi at \SI{400}{\degreeCelsius}. These findings are summarized in Fig.~\hyperref[fig:xrd_Mn5Ge3]{4(a)}. When Mn and Ge were co-deposited on a \ce{CrSi2} template (created through either Si-rich deposition or Cr diffusion into the substrate above \SI{350}{\degreeCelsius}), no crystallization was observed in RHEED up to temperatures of approximately \SI{485}{\degreeCelsius}.

We also grew selected \mbox{Mn-Ge/\SIdash{2}{\QL} CrSi/Si(111)} samples with 10-15\% excess Mn on the \SIdash{2}{\QL} CrSi template layer. As seen in Fig.~\hyperref[fig:xrd_Mn5Ge3]{4(b)}, two-stage annealing to \SI{250}{\degreeCelsius} resulted in a mixture of B20 MnGe and D8\textsubscript{8} \ce{Mn5Ge3} (space group No. 193 $P6_3/mcm$). Additional \textit{ex situ} annealing for 60 min after deposition of a \SI{20}{\nano\metre} amorphous Ge overlayer is able to completely convert the Mn-rich impurity \ce{Mn5Ge3} into MnGe. Annealing these films to \SI{375}{\degreeCelsius} converted some of the excess Mn to B20 MnSi, presumably due to Mn diffusion through the template layer and into the substrate. Interestingly, the annealing improved the roughness of the \ce{Mn5Ge3} phase, as indicated by the emergence of Laue oscillations around the \ce{Mn5Ge3}(0002) peak. Mn concentrations as large as \ce{Mn_{5.0}Ge} deposited on the \SIdash{2}{\QL} CrSi template resulted in the formation of (0001)-oriented \ce{Mn5Ge3} rather than other Mn-rich polytypes. In the case of Mn-Ge films with excess Ge, we found that the \mbox{Mn-Ge/\SIdash{2}{\QL} CrSi/Si(111)} films are heterogeneous with a mixture of B20 MnGe(111) and Ge(111).

\section{Structure and Strain Characterization}

\begin{figure*}[ht!]
	\centering
	\includegraphics[scale=1]{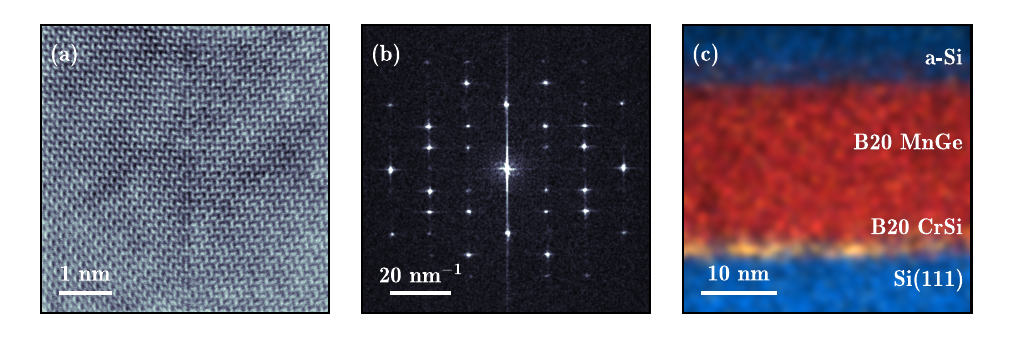}
	\caption{(a) HAADF-STEM image of a \SIdash{22.9}{\nano\metre} MnGe film depicting two chiral domains. (b) SADP calculated from the FFT of a lower magnification image of the domains in (a) encompassing the entire film thickness. (c) EDS elemental mapping of Mn, Ge, Cr, and Si. The substrate and capping layer are blue, and the MnGe film is shown in red. The orange layer at the Si-MnGe interface is the \SIdash{2}{\QL} CrSi template.}
	\label{fig:tem_crosssection}
\end{figure*}

To gain further insight into the nanoscale film and interface structure, we performed cross-sectional scanning transmission electron microscopy (STEM) imaging. A thin cross-sectional lamella was prepared by focused ion beam using a Thermo Fisher Scientific Helios 5 UX DualBeam FIB-SEM and imaged in a double-aberration corrected Thermo Fisher Scientific Spectra 300 microscope operated at \SI{200}{\kilo\volt}. Figure~\ref{fig:tem_crosssection} shows two high-angle annular dark-field (HAADF) images of a \SIdash{22.9}{\nano\metre}-thick MnGe film atop a \SIdash{2}{\QL} CrSi template with an \SIdash{8}{\nano\metre} Si capping layer. The electron beam is parallel to Si$[11\bar{2}]$. The well-known \emph{staircase} pattern observed in Fig.~\hyperref[fig:tem_crosssection]{5(a)} shows two domains with opposite stacking directions corresponding to opposite crystal chiralities. The symmetric selected area diffraction pattern (SADP) in Fig.~\hyperref[fig:tem_crosssection]{5(b)} obtained by fast Fourier transform (FFT) over the entire film thickness is composed of two asymmetric diffraction patterns, one from each chiral domain. A wider view of the whole film is presented in Fig.~\hyperref[fig:tem_crosssection]{5(c)}. The false color overlay is an elemental mapping of Mn, Ge, Cr and Si using energy-dispersive x-ray spectroscopy (EDS). The interfaces above and below the MnGe layer are sharp, and there is a distinct layer containing Cr at the film-substrate interface.

\begin{figure}[!htb]
	\centering
	\includegraphics[scale=1]{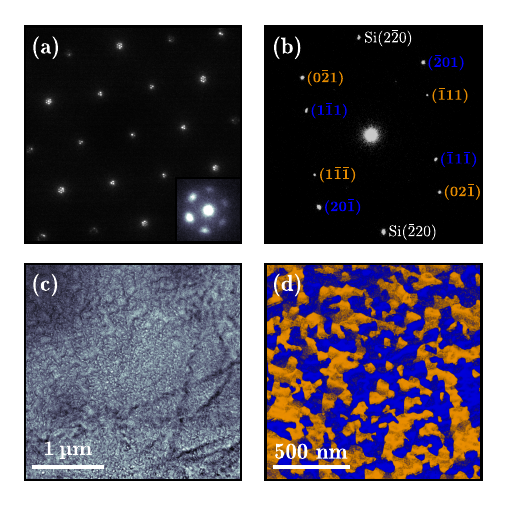}
	\caption{(a) Representative TEM SADP along the MnGe[111] zone axis for a \SIdash{29.4}{\nano\metre} MnGe film. The inset depicts the fine structure of the diffraction spots at higher magnification. (b) SADP of the MnGe[321] zone axis. (c) Bright-field image along the MnGe[111] zone axis. (d) False color dark-field image using the $(0\bar{2}1)$ spot (orange) and the ($1\bar{1}1$) spot (blue) in (b) which reveals the chiral domains.}
	\label{fig:tem_planview}
\end{figure}

To further characterize the film structure, we performed a detailed investigation of reciprocal space using TEM and XRD. Samples for plan-view TEM were prepared by low-angle mechanical polishing and attached to TEM grids with silver epoxy as described in Ref.~\cite{Robertson_2006}. Diffraction and imaging were done using the FEI Tecnai G2 F20 microscope. A sample SADP of the [111] zone axis of a \SIdash{29.4}{\nano\metre} MnGe film is shown in Fig.~\hyperref[fig:tem_planview]{6(a)}. The fine structure of each spot is due to double diffraction from the film and substrate~\cite{Karhu_2010}. We extracted the in-plane lattice spacings using the \textsc{fiji} software~\cite{Fiji}. The corresponding bright-field image in Fig.~\hyperref[fig:tem_planview]{6(c)} appears uniform except for scratches due to the polishing, and free of impurity precipitates, similar to the AFM in Fig.~\ref{fig:rheed_afm}.

To visualize the chiral domains seen in the HAADF-STEM images, we tilted the sample by approximately \ang{22.2} about Si$[2\bar{2}0]$ to reach the MnGe$[321]$ zone axis. Dark field imaging of the $(1\bar{1}0)$ and $(0\bar{2}1)$ spots in Fig.~\hyperref[fig:tem_planview]{6(b)} revealed domains with sizes of approximately 100-\SI{800}{\nano\metre}, like those in Fig.~\hyperref[fig:tem_planview]{6(d)} across MnGe samples with thicknesses ranging from 4 to \SI{40}{\nano\metre}. There is no preferential crystallographic orientation to the domain boundaries.

\begin{figure}[!htb]
	\centering
	\includegraphics[scale=1]{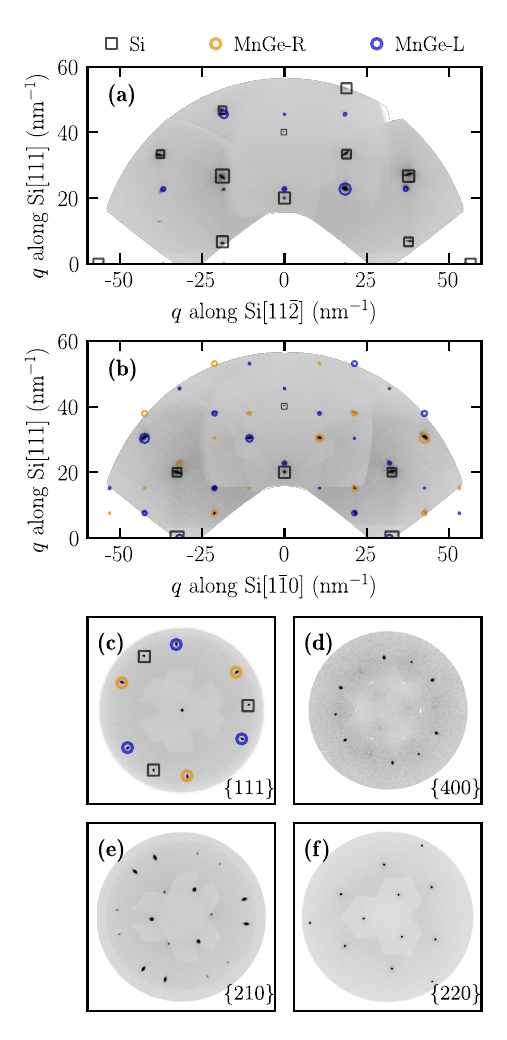}
	\caption{Slices through the XRD-RSM along the (a) Si$[11\bar{2}]$ and (b) Si$[1\bar{1}0]$ in-plane directions. Calculated peak positions for right- and left-handed rhombohedrally distorted MnGe are indicated by circles, along with squares for the substrate peaks. The size of the markers is proportional to the calculated intensity. (c), (d) Pole figures centered around (111) for MnGe and Si \{111\} and \{400\} families showing that MnGe spots occur $\pm\ang{30}$ from the Si spots. (e), (f) Pole figures centered around (111) for MnGe \{210\} and \{220\} families.}
	\label{fig:xrd_rsm}
\end{figure}

We performed a more comprehensive investigation of reciprocal space by collecting x-ray reciprocal space maps (XRD-RSMs) of the films using a Bruker D8 Venture diffractometer with Cu K$\alpha$ radiation. We combined three $\omega$ scans and one $\phi$ scan to cover a maximum scattering vector of approximately \SI{56.3}{\per\nano\metre} and visualized the 3D dataset using the \textsc{max3d} software~\cite{max3d}. We used the diffraction spots from the Si substrate as an internal calibration and refined the film spots using the Bruker \textsc{apex} software and \textsc{xrayutilities} code~\cite{xrayutilities}. No peaks were found that could not be attributed to the substrate or B20 MnGe film. We refined two unit cells, beginning first with B20 MnGe. While this model roughly described the data, it did not accurately capture the peak locations across all of reciprocal space. The second model was the rhombohedral distortion of the B20 structure in Fig.~\hyperref[fig:MnGe_structure]{1(b)} which gave significantly better agreement with the peak positions in the data, especially at larger $Q$. In Fig.~\hyperref[fig:xrd_rsm]{7(a)} and \hyperref[fig:xrd_rsm]{7(b)}, we present slices through the in-plane high-symmetry directions of the RSM for the same \SIdash{22.9}{\nano\metre} MnGe film in Fig.~\ref{fig:tem_crosssection}.

The calculated peak positions for rhombohedrally distorted B20 MnGe are annotated in Fig.~\ref{fig:xrd_rsm} together with the Si (111) substrate peaks. Along the Si$[1\bar{1}0]$ direction, the RSM is symmetric and agrees with the SADP in Fig.~\hyperref[fig:tem_crosssection]{5(b)}, which is only explained by the superposition of two RSMs from both right- and left-handed MnGe. Figures~\hyperref[fig:xrd_rsm]{7(c)}--\hyperref[fig:xrd_rsm]{7(f)} present (111)-centered pole figures extracted from the XRD-RSMs. The \{111\} and \{400\} pole figures show that MnGe spots occur with a \ang{30} rotation relative to the substrate. The six MnGe spots correspond to three spots from two chiralities, like what was observed with SANS in Ref.~\cite{Kanazawa_2017}.

\begin{figure}[!htb]
	\centering
	\includegraphics[scale=1]{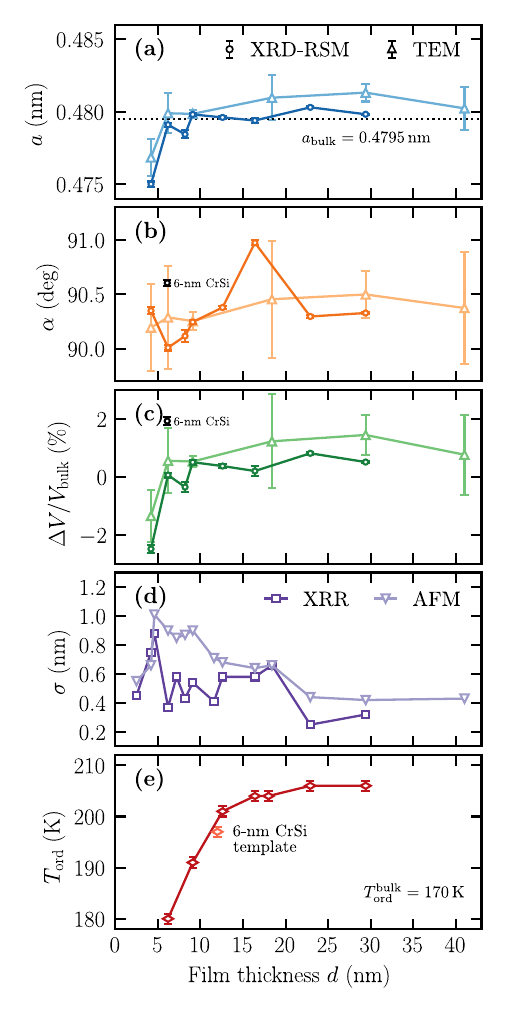}
	\caption{(a) Rhombohedral lattice constant and (b) rhombohedral angle of MnGe vs film thickness, calculated from XRD-RSMs and TEM. The bulk lattice parameter of undistorted MnGe is noted ($\alpha=\ang{90}$). (c) The volume strain $\Delta V/V_{\mathrm{bulk}}$ of the films. The rhombohedral angle and volume strain of the \SIdash{6}{\nano\metre} CrSi sample are also noted. (d) Roughness of the MnGe layer vs film thickness, obtained by XRR and AFM. The XRR measures the roughness of the MnGe layer and AFM measures the roughness of the film surface. (e) Ordering temperature vs thickness, estimated from the maximum in the dc susceptibility at low fields ($\mu_0 H \leq \SI{100}{\milli\tesla}$). All MnGe films in this figure were grown with a \SIdash{2}{\QL} CrSi template, except one sample noted in (e) which used a \SIdash{6}{\nano\metre} template.}
	\label{fig:params_vs_d}
\end{figure}

The lattice constant $a$ and rhombohedral angle $\alpha$ of the distorted B20 films are given in Fig.~\hyperref[fig:params_vs_d]{8(a)} and \hyperref[fig:params_vs_d]{8(b)}, calculated using both the TEM plane spacings and XRD-RSM refinement. All films studied exhibited $\alpha>\ang{90}$, similar to MnSi~\cite{Karhu_2012} and \ce{Fe_{1-x}Co_xSi}~\cite{Porter_2012} on Si(111). We also show the volume strain $\Delta V / V_{\textrm{bulk}}$ in Fig.~\hyperref[fig:params_vs_d]{8(c)}, which shows that the unit cell is larger than the bulk value for MnGe thicknesses above \SI{9}{\nano\metre}, but there is a contraction of the unit cell for thinner films. We also note that the TEM and XRD-RSM data generally agree, but the uncertainties in parameters obtained from the XRD-RSM are much lower.

\section{Magnetic Characterization with out-of-plane field}
The magnetic moment of the films was measured using the MPMS-XL SQUID magnetometer with an applied field oriented parallel to the MnGe[111] direction. This field direction is expected to be along the helical wavevector for MnGe films~\cite{Kanazawa_2017}. The diamagnetic contribution of the Si substrate $\chi_{\rm{Si}}$ is typically obtained from the high-field portion of the $M$ vs $H$ loops, far above the saturation field. For the case of our films, this is not possible for two reasons. First, the high-field susceptibility of bulk MnGe is small, but nonzero ($\chi_{\rm{HF}}\approx\SI{4}{\kilo\ampere\per\metre\per\tesla}$ at \SI{100}{\kelvin})~\cite{Kanazawa_2011}. Our previous work showed that MnSi films exhibit a similar $\chi_{\rm{HF}}$ to bulk~\cite{Meynell_2014} and we expect this to hold for MnGe.

Second, the saturation field of our films is much larger than the maximum applied field of the magnetometer ($\pm$\SI{7}{\tesla}) at low temperatures. We therefore neglect the small temperature dependence of the low-temperature susceptibility and estimate $\chi_{\rm{Si}}$ from $M$ vs $H$ loops at temperatures where the film is saturated. For each $M$ vs $H$ loop, the temperature was stabilized at $+\SI{7}{\tesla}$. The resultant loops were reproducibly symmetric in field, and only the positive field branches are shown.

\begin{figure}[!htb]
	\centering
	\includegraphics[scale=1]{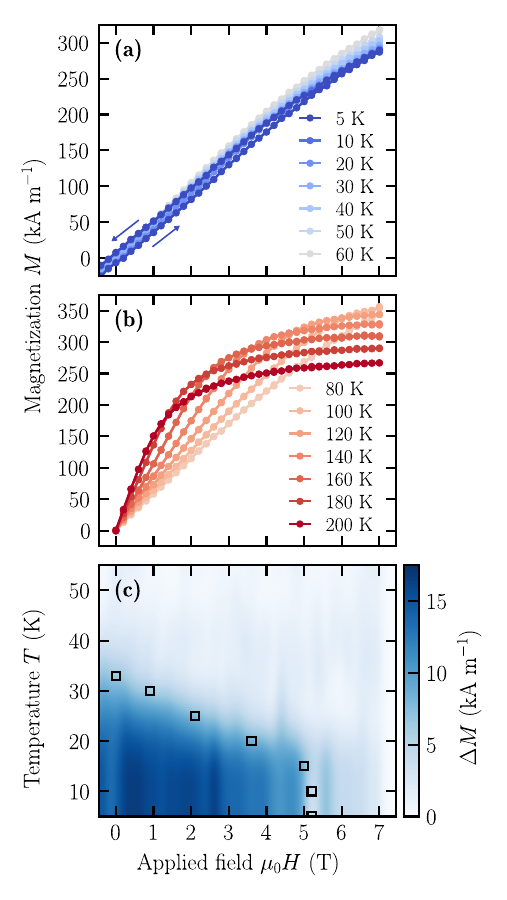}
	\caption{(a), (b) Corrected $M$ vs $H$ loops for the \SIdash{22.9}{\nano\metre}-thick film with field applied along MnGe[111]. The only feature is the second-order transition between conical and saturated states. (c) Colormap showing the difference $\Delta M$ between decreasing and increasing magnetization branches. The squares indicate the field-temperature points where hysteresis disappears.}
	\label{fig:squid_loops}
\end{figure}

Representative $M$ vs $H$ loops for the \SIdash{22.9}{\nano\metre} thick MnGe film are given in Fig.~\hyperref[fig:squid_loops]{9(a)} and \hyperref[fig:squid_loops]{9(b)}. At higher temperatures ($T>\SI{60}{\kelvin}$), the magnetization increases linearly before saturating without hysteresis. For these temperatures, the shape of the $M$ vs $H$ data is the same as observed in the conical phase of other B20 films in out-of-plane fields. Neutron scattering in both MnGe(111) films~\cite{Kanazawa_2017} as well as MnSi(111) films~\cite{Karhu_2011} confirm the conical magnetic order.

At low temperatures, (below approximately \SI{35}{\kelvin}), there is a clear difference between increasing and decreasing magnetization branches. The largest difference between magnetization branches \mbox{$\Delta M = M_{\mathrm{dec}} - M_{\mathrm{inc}} \approx \SI{16.5}{\kilo\ampere\per\metre}$} occurs at \SI{5}{\kelvin}. Figure~\hyperref[fig:squid_loops]{9(c)} depicts $\Delta M$ and shows the temperatures and fields where it disappears. This hysteresis is not associated with a simple conical state, since the cone angle simply varies continuously with a field applied along $Q$. The remanent moment and $\Delta M$ are also too large to be explained by a small in-plane component of the applied field due to sample misalignment inside the magnetometer. The only feature in the $M$ vs $H$ loops is the second-order transition from the conical state into the saturated state. This transition occurs at a field $H_{\mathrm{c}_2}$ that is defined by the minimum in $\dv*[2]{M}{H}$.

\begin{figure}[!htb]
	\centering
	\includegraphics[scale=1]{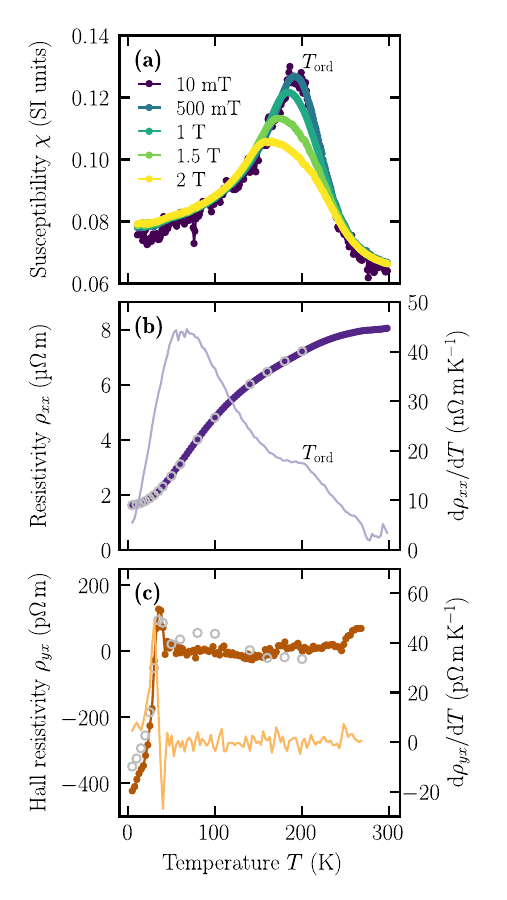}
	\caption{(a) Calculated dc susceptibility vs temperature for a \SIdash{9.1}{\nano\metre}-MnGe film with $H\parallel[111]$. The ordering temperature is estimated by the position of the peak at low applied field. (b) Zero-field longitudinal and (c) transverse resistivity for the \SIdash{22.9}{\nano\metre} MnGe film. The secondary axes (lighter color) show the differential resistivity. The circles in (b) and (c) are values extracted from the $\rho$ vs $H$ measurements.}
	\label{fig:M_rho_vs_T}
\end{figure}

The dc susceptibility $\chi=\dv*{M}{H}$ for constant applied fields is given in Fig.~\hyperref[fig:M_rho_vs_T]{10(a)}. The peak in the susceptibility at \SI{10}{\milli\tesla} provides an estimate of the ordering temperature $T_{\mathrm{ord}}$ in the material. As the field increases, the peak shifts to lower temperatures and corresponds to a transition from a conical state into a field-polarized ferromagnetic state. The thickness dependence of $T_{\mathrm{ord}}$ is shown in Fig.~\hyperref[fig:params_vs_d]{8(e)}. There is an enhancement of $T_{\mathrm{ord}}$ relative to bulk for all thicknesses presented here. The largest ordering temperature of \SI{206}{\kelvin} was observed for MnGe thicknesses of \SI{22.9}{\nano\metre} and \SI{29.4}{\nano\metre}. It has been reported that a \SIdash{160}{\nano\metre} MnGe film on a \SIdash{2}{\nano\metre} MnSi template layer also had an ordering temperature of \SI{206}{\kelvin}, and there was still an enhancement from bulk at \SI{1.8}{\micro\metre} MnGe thickness~\cite{Kanazawa_2017}.

We also conducted transport measurements using a DynaCool PPMS with maximum field of \SI{14}{\tesla}. The samples were cleaved into rectangles (nominally ${\SI{6.0}{\milli\metre}\times\SI{1.5}{\milli\metre}}$) and constant-current and transverse voltage leads were wire bonded to the surface. It is important to note that CrSi has metallic conductivity~\cite{Banik_2020} and that the parallel resistivity of the film and \SIdash{2}{\QL} CrSi template are being measured. Since the MnGe films are significantly thicker than the CrSi layer, the transport signal should be dominated by MnGe.

The longitudinal $\rho_{xx}$ and transverse resistivity $\rho_{yx}$ were calculated using the field-symmetrization and antisymmetrization requirements, respectively. Figure~\hyperref[fig:M_rho_vs_T]{10(b)} and \hyperref[fig:M_rho_vs_T]{10(c)} show the temperature dependence of $\rho_{xx}$ and $\rho_{yx}$ at zero field. We applied the symmetrization and antisymmetrization requirements to zero-field warmed data, which were collected after field cooling in $\pm\SI{14}{\tesla}$. Due to the complexity of symmetrizing temperature scans compared to field scans, we include values extracted from the $\rho$ vs $H$ measurements in the figure as a reference. There are features at low temperatures which are not observed in the magnetometry, including a reversal in the sign of the Hall resistivity at \SI{31}{\kelvin}. The $T_{\mathrm{ord}}$ estimated from features in the transport data is in agreement with the magnetometry.

\begin{figure*}[!htb]
	\centering
	\includegraphics[scale=1]{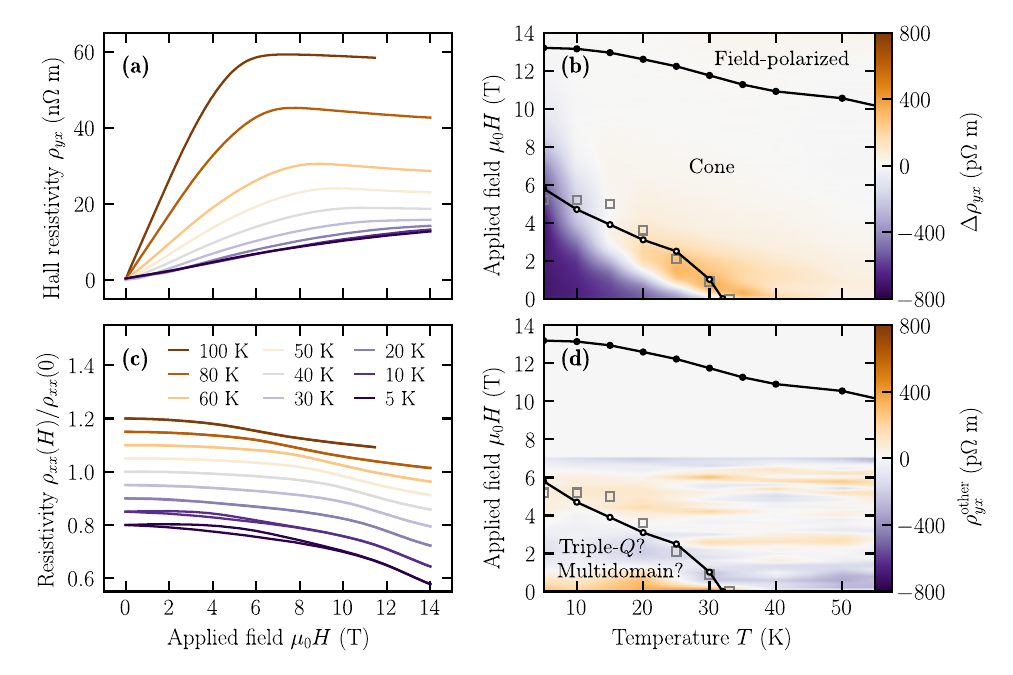}
	\caption{(a) Transverse and (c) normalized longitudinal resistivity vs field for the \SIdash{22.9}{\nano\metre}-thick MnGe film. (b), (d) Phase diagram for this sample with $H\parallel[111]$. The circles are features derived from the resistivity and the squares are points where the hysteresis disappears in the $M$ vs $H$ data. The colormap in (b) is the difference between decreasing and increasing field branches of $\rho_{yx}$ and the colormap in (d) is the unaccounted $\rho_{yx}^{\mathrm{other}}$ contribution to the resistivity after fitting Eq.~\ref{eq:hall_resistivity} to the data. The closed (open) circle phase boundaries are $H_{\mathrm{c}_2}$ ($H_{\mathrm{c}_1}$).}
	\label{fig:resistivity_all}
\end{figure*}

The isothermal field dependence of $\rho_{yx}$ and $\rho_{xx}$ are shown in Fig.~\hyperref[fig:resistivity_all]{11(a)} and \hyperref[fig:resistivity_all]{11(c)}. The saturation field $H_{\mathrm{c}_2}$ is extracted from these data by examining features in the derivatives, as was done for the magnetization. There are additional features in the resistivity that are not observed in the magnetometry. There is a deviation from a field-linear behavior in $\rho_{yx}$ which is accompanied by a crossing of the decreasing and increasing field branches: a sign reversal in $\Delta\rho_{yx} = \rho_{yx}^{\mathrm{dec}}(H) - \rho_{yx}^{\mathrm{inc}}(H)$. One of the more conspicuous features is a maximum in $\dv*{\rho_{yx}}{H}$ below approximately \SI{40}{\kelvin} which is also observed in $\dv*{\rho_{yx}}{T}$ at zero field [cf. Fig.~\hyperref[fig:M_rho_vs_T]{10(c)}]. We denote the fields of this feature as $H_{\mathrm{c}_1}$.

To address these anomalies and the topological nature of any low-temperature phases in MnGe films, we fitted the transverse resistivity $\rho_{yx}$ to the model from Ref.~\cite{Meynell_2014_Hall} which accounts for ordinary and anomalous Hall effects. The Hall resistivity in this model is
\begin{equation}
	\rho_{yx} = R_0 \mu_0H + A M \rho_{xx} + B M {\rho_{xx}}^2 + \rho_{yx}^{\rm{other}}\,,
	\label{eq:hall_resistivity}
\end{equation}
where $R_0 \mu_0H$ is the ordinary Hall effect, $A M \rho_{xx}$ accounts for extrinsic skew scattering, and $B M {\rho_{xx}}^2$ is a combined term describing extrinsic side-jump scattering and intrinsic contributions to the Hall resistivity. Due to the nonzero susceptibility above the saturation field and the inability to saturate the MnGe films at low temperatures, the Hall coefficient $R_0$ could not be determined prior to the fitting, as is commonly performed. Instead, we treated $R_0$ as a fitting parameter along with $A$ and $B$ (see the Supplemental Material for these fitting parameters and additional information regarding the transport measurements~\cite{supp}). Alternatively, one study substituted the $R_0$ from a bulk polycrystal when fitting the Hall conductivity in MnGe films~\cite{Hayashi_2021}. The $\rho_{yx}^{\rm{other}}$ contribution is often attributed to a topological Hall effect $\rho_{yx}^{\rm{T}}$ due to the Berry phase that the electron acquires while following a magnetization texture with nontrivial topology such as a spin hedgehog~\cite{Kanazawa_2011, Kanazawa_2012, Tanigaki_2015, Kanazawa_2016, Kanazawa_2017, Kanazawa_2020, Fujishiro_2021, Kitaori_2021, Hayashi_2021, Repicky_2021}. Additional contributions to the Hall effect in $\rho_{yx}^{\rm{other}}$ can arise from the helical structure~\cite{Lux_2020}, as found in MnSi films~\cite{Meynell_2014_Hall}. However, no evidence of such a contribution was found in MnGe films reported here.

The colormaps in Fig.~\hyperref[fig:resistivity_all]{11(b)} and \hyperref[fig:resistivity_all]{11(d)} depict $\Delta\rho_{yx}$ and $\rho_{yx}^{\rm{other}}$, along with the fields $H_{\mathrm{c}_1}$ and $H_{\mathrm{c}_2}$. The $\Delta M$ identified in Fig.~\hyperref[fig:squid_loops]{9(c)} are also shown, and align well with the $H_{\mathrm{c}_1}$ boundary. The phase diagram therefore appears to consist of three regions: a field-polarized state above $H_{\mathrm{c}_2}$, a conical state between $H_{\mathrm{c}_1}$ and $H_{\mathrm{c}_2}$, and an additional low-temperature phase below $H_{\mathrm{c}_1}$. The significant hysteresis in the Hall measurements, signaled by nonzero $\Delta \rho_{yx}$ is largely confined to this region. The corresponding $\rho_{yx}^{\rm{other}}$ shown in Fig.~\hyperref[fig:resistivity_all]{11(d)} extracted from Eq.~\ref{eq:hall_resistivity} does not show any significant contribution.

\section{Discussion}
Though it is likely not feasible to confirm the structure of the \SIdash{2}{\QL} CrSi template layers using XRD, the observed RHEED is sufficient to conclude that the amorphous {$\SIdash{0.5}{\QL}$ Si/$\SIdash{1}{\QL}$ Cr/$\SIdash{0.5}{\QL}$ Si} Higashi trilayer crystallizes upon annealing. While B20 CrSi and C40 \ce{CrSi2} may exhibit similar RHEED patterns, we note that the apparent drop in quality of the RHEED pattern with increasing temperature signals a structural change in the film. We suggest that this is due to diffusion of Si into the template which forms \ce{CrSi2}. As suggested in Refs.~\cite{Wetzel_1987, Wetzel_1988, Mahan_1993}, CrSi/Si(111) is converted to \ce{CrSi2} accompanied by poor epitaxy which explains the appearance of transmission spots in the RHEED pattern. The fact that we were only able to grow MnGe or CrSi films atop the \SIdash{2}{\QL} CrSi is strong evidence that the template is indeed B20 CrSi, rather than another Cr-Si polymorph. We also wish to note that two recent studies reported the growth of (210)-textured monoclinic and cubic CrSi on Si(111) and claimed that the films exhibited ferromagnetic order between 3 and \SI{300}{\kelvin} including a moment of approximately 1$\mu_{\mathrm{B}}$ per Cr atom at \SI{300}{\kelvin}~\cite{Galkin_2025_1, Galkin_2025_2}. We do not observe this behavior (see the Supplemental Material for characterization of the \SIdash{6}{\nano\metre} CrSi sample~\cite{supp}).

The data presented in Figs.~\ref{fig:xrd_MnGe} and \ref{fig:xrd_rsm} show that B20 MnGe can be stabilized atop a template of B20 CrSi. To date, MnSi and FeGe have also been used. So long as the annealing temperature is kept below \SI{250}{\degreeCelsius}, diffusion of Mn into the Si substrate appears minimal, based on XRD and TEM. This template is also capable of nucleating epitaxial B20 CrSi (while we found that a solid phase reaction could not reliably do so), and even D8$_{8}$ \ce{Mn5Ge3}. This invites the possibility of using varying thicknesses of a CrSi buffer layer to engineer strain in the overlaying film without impacting its magnetic properties. In particular, we note that the lattice mismatch for FeGe/Si(111) is $-0.31\%$ whereas it is $+1.28\%$ for FeGe/CrSi if the CrSi is fully relaxed.

Since the metastable MnGe structure does not appear to crystallize atop any template other than B20 MnSi, FeGe or CrSi, we hypothesize that during the postdeposition annealing step, crystallization initiates at the CrSi interface and proceeds upwards. Combining the RHEED, HAADF-STEM, and XRD-RSM data, we can conclude the following epitaxial relationship for our films:
\begin{align*}
	&\langle 1\bar{2}1 \rangle\ce{MnGe}(111) \parallel \langle1\bar{2}1\rangle\ce{CrSi}(111) \parallel [1\bar{1}0]\ce{Si}(111).
\end{align*}
The chirality of the MnGe domains is likely dictated by the underlying CrSi template. We observed evidence for twin domains in our film, which are visualized directly in the opposite stacking directions of MnGe$[1\bar{2}1]$ planes in Fig.~\hyperref[fig:tem_crosssection]{5(a)} and by dark-field imaging on the MnGe$[321]$ zone axis in Fig.~\hyperref[fig:tem_planview]{6(d)}. Our XRD-RSM measurements report similar findings to what Ref.~\cite{Trabel_2017} stated for MnSi, where it is clear that these twin domains are rotated by $\pm\ang{30}$ relative to the Si substrate and that the asymmetric intensities of the $\{210\}$ family in Fig.~\hyperref[fig:xrd_rsm]{7(e)} indicate that there cannot be any rotational twinning (see the Supplemental Material for a calculation of pole figure intensities for the four morphologies~\cite{supp}).

The calculated intensities of the $\{210\}$ reflections for the four stacking morphologies dictate that twins related through an inversion operation must occur together: \mbox{MnGe-$R[111]$} + \mbox{MnGe-$L[\bar{1}\bar{1}\bar{1}]$} or \mbox{MnGe-$L[111]$} + \mbox{MnGe-$R[\bar{1}\bar{1}\bar{1}]$}. For example, the pair \mbox{MnGe-$R[111]$} + \mbox{MnGe-$L[111]$} does not reproduce the observed pattern. In principle, another similar analysis of another intensity-split reflection would allow us to distinguish which of these pairs is present in our films. However, the intensity asymmetry is too small to detect within the resolution of our measurements, nor can we elucidate the exact stacking manner within the resolution of our HAADF-STEM measurements. The left and right \textit{staircase} stacking sequences seen in Fig.~\hyperref[fig:tem_crosssection]{5(a)} are explained by both sets of chiral pairs, and the individual atoms within $\SI{1}{\QL}$ must be resolved in order to distinguish which is present. In addition to structural intensity asymmetry, an additional asymmetry is introduced by anomalous dispersion which is amplified near an x-ray absorption resonance. We note that Ref.~\cite{Morikawa_2020} reported that \mbox{MnSi-$L[111]$} + \mbox{MnSi-$R[\bar{1}\bar{1}\bar{1}]$} was favored in MnSi/Si(111) using detailed reflectivity measurements.

The fact that we report a rhombohedral angle \mbox{$\alpha>\ang{90}$} for all MnGe films is unexpected since the $+1.91\%$ epitaxial mismatch between MnGe and Si suggests compressive strain and therefore $\alpha<\ang{90}$. Strain data presented in Ref.~\cite{Kanazawa_2017} for MnGe/\SIdash{2}{\nano\metre} MnSi are consistent with this unexpected behavior although no discussion about its origin was given. While both MnSi~\cite{Karhu_2012} and \ce{Fe_{1-x}Co_xSi}~\cite{Porter_2012} films on Si(111) also have a rhombohedral structure with $\alpha > 90$, this is expected based on their negative lattice mismatch. However, in the case of MnGe/Si(111) it is not clear why the positive lattice mismatch does not yield an in-plane compressive strain. In the case of MnSi/Si(111), Ref.~\cite{Figueroa_2016} reports extended x-ray absorption fine-structure measurements that reveal larger relative changes to the positions of the Si atomic sites than to the unit cell itself. There are large differences in the coordination number between Si and MnSi. These differences could be driving changes in the bond angles near the MnSi/Si interface that are driving the rhombohedral distortion. This may also be true of the MnGe/CrSi/Si interface. 

The enhancement of $T_{\mathrm{ord}}$ relative to bulk persists for thick films that are nominally strain-relaxed. A similar behavior has been reported for MnSi films~\cite{Karhu_2010, Figueroa_2016} which suggests a far-reaching influence of the Si substrate on epitaxial B20 films, possibly driven by changes within the unit cell that are not evident in XRD strain analysis.

We also note that the MnSi/SiC(0001) system is also expected to possess an in-plane compressive strain, but a tensile strain was found instead~\cite{Meynell_2016}. This was attributed to a dewetting of the film from the substrate and a thermal residual in-plane tensile strain after annealing to \SI{500}{\degreeCelsius}. Perhaps changes to MnSi bond angles near the MnSi/SiC interface are also influencing the strain in this system as well.

At temperatures above \SI{60}{\kelvin}, the linear dependence of the $M$ vs $H$ loops below saturation and the absence of $\rho_{yx}^{\rm{other}}$ in fit of the Hall-effect data are consistent with the presence of a conical phase, as reported by others~\cite{Kanazawa_2017, Fujishiro_2021, Hayashi_2021}. There is some evidence of $\rho_{yx}^{\rm{other}}$ below \SI{35}{\kelvin}, although the additional contribution to the Hall effect in the \SIdash{22.9}{\nano\metre} MnGe film is only $\sim\SI{0.1}{\nano\ohm\metre}$ compared to $\sim\SI{1}{\nano\ohm\metre}$ in the thicker 80--\SIdash{300}{\nano\metre} films~\cite{Kanazawa_2017, Fujishiro_2021, Hayashi_2021}. It appears that the characteristic large negative contribution to $\rho_{yx}$ is thickness dependent, since this signal is diminished for the {\SIdash{50}{\nano\metre} MnGe/\SIdash{5}{\nano\metre} FeGe/Si(111)} film in Ref.~\cite{Repicky_2021} and largest for the bulk like \SI{3}{\micro\metre} films~\cite{Kanazawa_2020, Kitaori_2021}. This reduction would be consistent with the interpretation given in Ref.~\cite{Kanazawa_2017} that the reorientation of helical wavevectors toward the film normal leads to a more dilute distribution of emergent monopoles in the hedgehog lattice.

Nevertheless, we identify a low-temperature phase boundary defined by the onset of hysteresis in the magnetization, which coincides with the feature $H_{\mathrm{c}_1}$ in the Hall-effect data. The location of the phase is consistent with the one defined by transport measurements in other films~\cite{Kanazawa_2017, Kanazawa_2020} where two twin threefold SANS diffraction patterns were observed. The field dependent hysteresis has not been previously reported in this system. The origin of this hysteresis is not known, nor do we know what role surfaces and magnetocrystalline anisotropy are playing. A uniaxial compression of bulk polycrystalline MnGe causes a reorientation of the helical wavevector along the direction of the applied stress \cite{Deutsch_2014}. The $\alpha > \ang{90}$ rhombohedral distortion is therefore expected to drive the helical wavevector toward the $[111]$ direction, as does the demagnetizing field, consistent with magnetometry at higher temperatures. As the temperature decreases, cubic anisotropy and anisotropic exchange would also be expected to increase in strength and contribute to a reorientation of the helical wavevector toward the $\langle 100 \rangle$ directions.

The presence of helical domains creates topological defects, as observed in scanning probe measurements of FeGe~\cite{Schoenherr_2018} and MnGe~\cite{Repicky_2021}. This means that the magnetic transition to a multi-domain state is first order. Hysteresis is observed in the helical reorientation in MnSi from [111] to [100] in an applied field~\cite{Bauer_2017}. Therefore, the onset of hysteresis reported in Fig.~\hyperref[fig:squid_loops]{9(c)} is consistent with the formation of a multidomain state. Neither the hedgehog state nor the multidomain state can be definitively identified from our existing data.

An additional possible contributing factor are finite-size effects that can produce a canting of the helical propagation vector away from the $[111]$ direction to form the oblique spirals~\cite{Leonov_2020} observed in micromagnetic calculations of chiral magnetic films, but not yet measured experimentally. 

\section{Summary}
We have demonstrated the growth of epitaxial MnGe(111) films on Si(111) using an ultrathin template of B20 CrSi. Contrary to those reported previously, the films investigated in this paper are free from the influence of a neighboring magnetic layer and have thicknesses comparable to the helical wavelength. A large portion of the magnetic phase diagram below the ordering temperature is a single-$Q$ conical state. We observed subtle hysteresis in the magnetization below a temperature of \SI{35}{\kelvin}, coincident with features in the Hall resistivity. These signals indicate the presence of an additional phase, which is not observed in bulk. However, our magnetometry and transport measurements are not able to distinguish between multidomain single-$Q$ states and triple-$Q$ spin-hedgehog states that are debated in the literature. A more direct probe such as neutron or resonant x-ray scattering is required to determine the magnetic structure of this low-temperature phase in the thin film limit.

\section*{ACKNOWLEDGMENTS}
We acknowledge the support of the Natural Sciences and Engineering Research Council of Canada (NSERC). We thank L. Kreplak at Dalhousie University for facilitating AFM measurements; V. Jarvis at MAX for assistance with x-ray RSM measurements; C. Lupien at Universit\'e de Sherbrooke for PPMS measurements; B. Ballesteros from ICN2 for assistance with TEM measurements; and M. D. Robertson at Acadia University for insightful discussions regarding TEM sample preparation and analysis.

We gratefully acknowledge the Science and Technology Facilities Council (STFC) for access to the ISIS Materials Characterisation Laboratory. This project has received funding from the European Union’s Horizon 2020 research and innovation programme under Grant Agreement No. 101007417 having benefited from the access provided by ICN2 in Bellaterra (Barcelona) within the framework of the NFFA-Europe Pilot Transnational Access Activity, proposal ID697. We acknowledge the use of instrumentation as well as the technical advice provided by the Joint Electron Microscopy Center at ALBA (JEMCA) and funding from Grant IU16-014206 (METCAM-FIB) to ICN2 funded by the European Union through the European Regional Development Fund (ERDF), with the support of the Ministry of Research and Universities, Generalitat de Catalunya.

\section*{DATA AVAILABILITY}
The data that support the findings of this article are openly available~\cite{data_repository}.

\end{document}


\title{Supplemental Material: Structure and magnetism of MnGe thin films grown with a nonmagnetic CrSi template}

\author{B.~D.~MacNeil}
\affiliation{Department of Physics and Atmospheric Science, Dalhousie University, Halifax, Nova Scotia, B3H 3J5, Canada}

\author{J.~S.~R.~McCoombs}
\affiliation{Department of Physics and Atmospheric Science, Dalhousie University, Halifax, Nova Scotia, B3H 3J5, Canada}

\author{D.~Kalliecharan}
\affiliation{Department of Physics and Atmospheric Science, Dalhousie University, Halifax, Nova Scotia, B3H 3J5, Canada}

\author{J.~Myra}
\affiliation{Department of Physics and Atmospheric Science, Dalhousie University, Halifax, Nova Scotia, B3H 3J5, Canada}

\author{M.~Pula}
\affiliation{Department of Physics and Astronomy, McMaster University, Hamilton, Ontario, L8P 4N3, Canada}

\author{J.~F.~Britten}
\affiliation{McMaster Analytical X-ray Diffraction Facility, McMaster University, Hamilton, Ontario, L8P 4N3, Canada}

\author{G.~B.~G.~Stenning}
\affiliation{ISIS Neutron and Muon Source, STFC Rutherford Appleton Laboratory, Didcot, OX11 0QX, United Kingdom}

\author{K.~Gupta}
\affiliation{Catalan Institute of Nanoscience and Nanotechnology (ICN2), Campus UAB, Bellaterra, Barcelona, 08193, Spain}

\author{G.~M.~Luke}
\affiliation{Department of Physics and Astronomy, McMaster University, Hamilton, Ontario, L8P 4N3, Canada}

\author{T.~L.~Monchesky}
\affiliation{Department of Physics and Atmospheric Science, Dalhousie University, Halifax, Nova Scotia, B3H 3J5, Canada}
%

\maketitle

\section*{RHEED images of CrSi and MnGe films}
\begin{figure*}[!htb]
	\centering
	\includegraphics[scale=1]{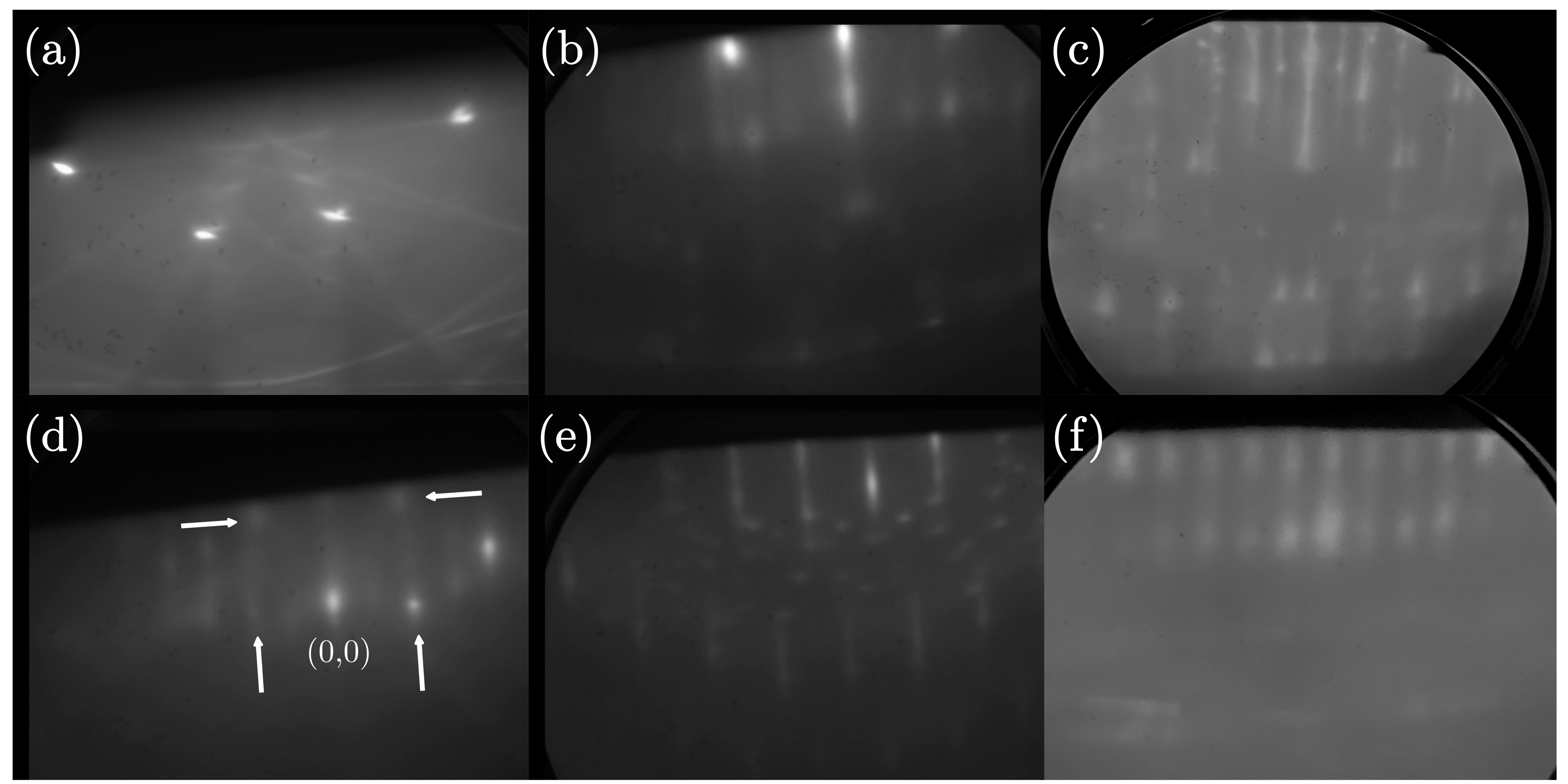}
	\caption{RHEED images of:
		(a) As-deposited Higashi CrSi trilayer at room temperature prior to annealing, viewed along Si$[1\bar{1}0]$. (b) Higashi CrSi trilayer that has been converted to \ce{CrSi2}, also viewed along Si$[1\bar{1}0]$. (c) A thick ($\sim\SIdash{16}{\nano\metre}$) \ce{CrSi2} film. (d) A surface partially covered by both CrSi and \ce{CrSi2} after excess annealing of the \SIdash{2}{\QL} CrSi template, viewed along CrSi$[1\bar{2}1]$. The arrows show the additional diffraction features which are ascribed to \ce{CrSi2}. The \ce{CrSi2} diffraction features [similar to those in (b)] appeared independent of wafer rotation. (e) Coexistence of transmission spots and $\surd 3 \times \surd 3$ pseudoreconstruction viewed along Si$[11\bar{2}]$ of a \SIdash{20}{\nano\metre} MnGe film grown on a mixed-phase template of \SIdash{2}{\QL} CrSi and \ce{CrSi2}. (f) \ce{Mn_{1.15}Ge}/\SIdash{2}{\QL} CrSi film which appeared as single-phase \ce{Mn5Ge3} in XRD.}
	\label{fig:rheed}
\end{figure*}

RHEED can be used as a diagnostic  tool to help identify the optimal annealing temperatures and correct stoichiometric ratios of the deposited precursor elements in order to form B20 films. In Fig.~\ref{fig:rheed}, we present selected RHEED images to illustrate its use for this purpose. The diffuse spots in Fig.~\ref{fig:rheed}(a) were observed along Si$[1\bar{1}0]$ after the Higashi trilayer~\cite{Higashi_2009} was deposited at room temperature. The lack of fractional order spots, and diffuse nature of the principal spots indicate that the layer has likely not crystallized and is not a suitable template for the growth of MnGe.

After annealing, this pattern is transformed to the $\surd 3 \times \surd 3$ pseudoreconstruction in Fig.~2(e) of the main text. When the \SIdash{2}{\QL} CrSi template is further annealed above \SI{420}{\degreeCelsius}, the reconstruction deteriorated to the pattern shown in Fig.~\ref{fig:rheed}(b), which signaled the formation of \ce{CrSi2}. This identification is accomplished by comparison with the RHEED pattern of a single-phase \ce{CrSi2} film in Fig.~\ref{fig:rheed}(c), whose structure was verified with XRD. Our RHEED investigations found that the \SIdash{2}{\QL} CrSi template and thicker CrSi films could be converted to \ce{CrSi2} upon annealing to \SI{420}{\degreeCelsius} and \SI{350}{\degreeCelsius} respectively, likely caused by diffusion of Si from the substrate.

Fig.~\ref{fig:rheed}(d) shows an intermediate film where both CrSi and \ce{CrSi2} are present. The Higashi trilayer was originally deposited and annealed to \SI{390}{\degreeCelsius}, which resulted in the pattern from Fig.~2(e) of the main text. This layer was then subsequently annealed to approximately \SI{450}{\degreeCelsius}. In the figure, there are two sets of spots: one from CrSi, which moves when the sample is azimuthally rotated, and another which is independent of rotation. In the figure, the arrows indicate the additional diffraction features. We ascribe this second pattern to \ce{CrSi2} based on its similarity to the pattern in Fig.~\ref{fig:rheed}(b). The fact that this was independent of sample rotation is consistent with the poor epitaxy reported for \ce{CrSi2} below \SI{1100}{\degreeCelsius}~\cite{Mahan_1993}. In Fig~\ref{fig:rheed}(e) we show a RHEED pattern of a \SIdash{20}{\nano\metre} MnGe film that was grown atop a mixed-phase template of CrSi and \ce{CrSi2}. The $\surd 3 \times \surd 3$ pseudoreconstruction of MnGe is visible, along with transmission spots. Since the RHEED beam images a large area of the sample surface relative to the crystallographic domain size, we suggest that the $\surd 3 \times \surd 3$ pseudoreconstruction is resulting from areas of the film with MnGe/\SIdash{2}{\QL} CrSi and the additional spots arise from regions with Mn-Ge/\ce{CrSi2}.
\clearpage

RHEED is also helpful in identifying the composition of the Mn-Ge films deposited on the CrSi template. We found that the stability of the B20 MnGe was strongly dependent on the Mn concentration. The pattern in Fig.~\ref{fig:rheed}(f) depicts a sample Mn-rich \ce{Mn_{1.15}Ge}/\SIdash{2}{\QL} CrSi film that exhibited only \ce{Mn5Ge3} peaks in XRD and no evidence of the B20 phase. With only 10--15\% excess Mn, the RHEED pattern changes from one like the $\surd 3 \times \surd 3$ pseudoreconstruction shown in Fig.~2(j) and 2(k) of the main text to the one shown in Fig.~\ref{fig:rheed}(f). \emph{Ex situ} XRD confirmed that the film was \ce{Mn5Ge3}(0001). For the films discussed in the main text which exhibited coexistence of MnGe and \ce{Mn5Ge3}, the RHEED pattern was generally a superposition of those in Figs.~2(j) and 2(k) of the main text, and Fig.~\ref{fig:rheed}(f).

\section*{X-ray reflectometry}
\begin{figure*}[!ht]
	\centering
	\includegraphics[scale=0.95]{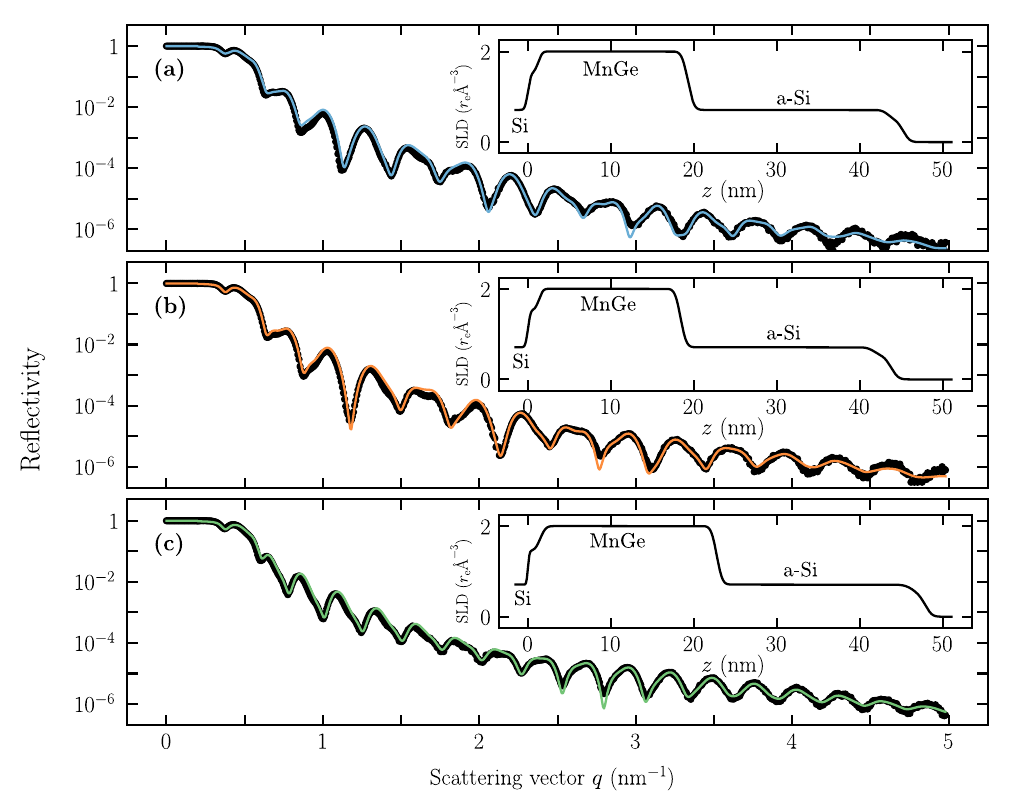}
	\caption{Sample XRR data of MnGe films grown with a \SIdash{2}{\QL} CrSi template, with thicknesses (a) \SI{18.0}{\nano\metre}, (b) \SI{17.0}{\nano\metre}, and (c) \SI{21.3}{\nano\metre}. The lines are fits obtained using \textsc{genx}. The insets show the SLD profiles obtained from these fits, which display interfacial roughnesses below \SI{0.8}{\nano\metre}.}
	\label{fig:xrr}
\end{figure*}

In Fig.~\ref{fig:xrr}, we show sample XRR measurements for three films. The data were fitted using the \textsc{genx} software~\cite{GenX}. Each fit included the Si substrate, CrSi template, MnGe film, Si capping layer, and \ce{SiO2}. The substrate roughness was in agreement with AFM. The CrSi template is difficult to resolve with XRR as only the first order Kiessig fringe is observable up to $q=\SI{5}{\per\nano\metre}$. The addition of CrSi into the model was necessary to achieve a satisfactory fit. Its presence is reflected in the extracted scattering length density (SLD) profiles. Layer thicknesses determined from, XRD, XRR, and HAADF-STEM were in agreement with one another and the nominal deposited values from the quartz-crystal microbalances.
 
\clearpage

\section*{X-ray reciprocal space mapping}
\begin{figure*}[!ht]
	\centering
	\includegraphics[scale=1]{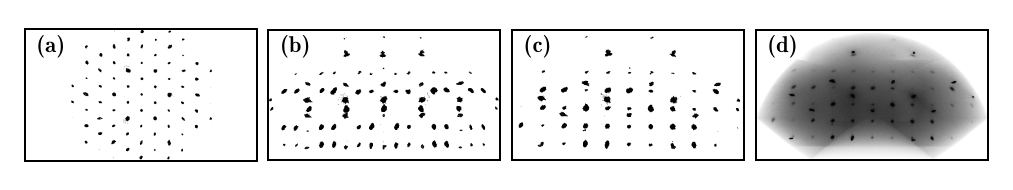}
	\caption{Illustration of the full XRD-RSM datasets visualized by \textsc{max3d}. (a) View of the Si$[1\bar{1}0]$--$[11\bar{2}]$ plane with Si$[111]$ out of the page. (b) View of the Si$[1\bar{1}0]$--$[111]$ plane with Si$[11\bar{2}]$ out of the page. (c) View of the Si$[11\bar{2}]$--$[111]$ plane with Si$[1\bar{1}0]$ out of the page. (d) The same view as in (c), with diffuse scattering to illustrate the mapped region of reciprocal space.}
	\label{fig:xrd-rsm}
\end{figure*}

We used the Bruker D8 Venture diffractometer to map reciprocal space for our films. We stitched three $\omega$ scans and one $\phi$ scan, as detailed in Table~\ref{tab:xrd_rsm_scans}. Additional \textsc{max3d}~\cite{max3d} visualizations of the XRD-RSMs are provided in Fig.~\ref{fig:xrd-rsm}.

\begin{table}[H]
	\centering
	\caption{Parameters of the four scans used to construct XRD-RSMs.}
	\label{tab:xrd_rsm_scans}
	\begin{ruledtabular}
		\begin{tabular}{ccccccc}
			&  & $2\theta$ & $\omega$ & $\phi$ & $\chi$ & \\
			\hline
			& Scan 1 & \ang{55} & \ang{5}--\ang{43} & \ang{0} & \ang{0} & \\
			& Scan 2 & \ang{55} & \ang{5}--\ang{43} & \ang{120} & \ang{0} & \\
			& Scan 3 & \ang{55} & \ang{5}--\ang{43} & \ang{240} & \ang{0} & \\
			& Scan 4 & \ang{55} & \ang{6} & \ang{0}--\ang{360} & \ang{35.26} & \\
		\end{tabular}
	\end{ruledtabular}
\end{table}

The noncentrosymmetric B20 crystal structure causes certain families of reflections to split into pairs with differing intensity. This effect is illustrated by the pole figure in Fig.~7(e) of the main text where the 24 spots from $\{210\}$ are split into two sets of 12 higher intensity and 12 lower intensity spots. In Table~\ref{tab:210_intensities}, we tabulate the intensities of the 24 $\{210\}$ reflections for the four morphologies, including corrections for anomalous dispersion and nuclear Thomson scattering. The values we calculate are in agreement with those obtained from \textsc{vesta}~\cite{Vesta} and \textsc{xrayutilities}~\cite{xrayutilities}. Note that we constructed the basis atoms in the B20 unit cell using $u_{\rm{Mn}}=0.136$ and $u_{\rm{Ge}}=0.846$ as obtained for a bulk polycrystal~\cite{Makarova_2012}.

\begin{figure*}[!htb]
	\centering
	\includegraphics[scale=1]{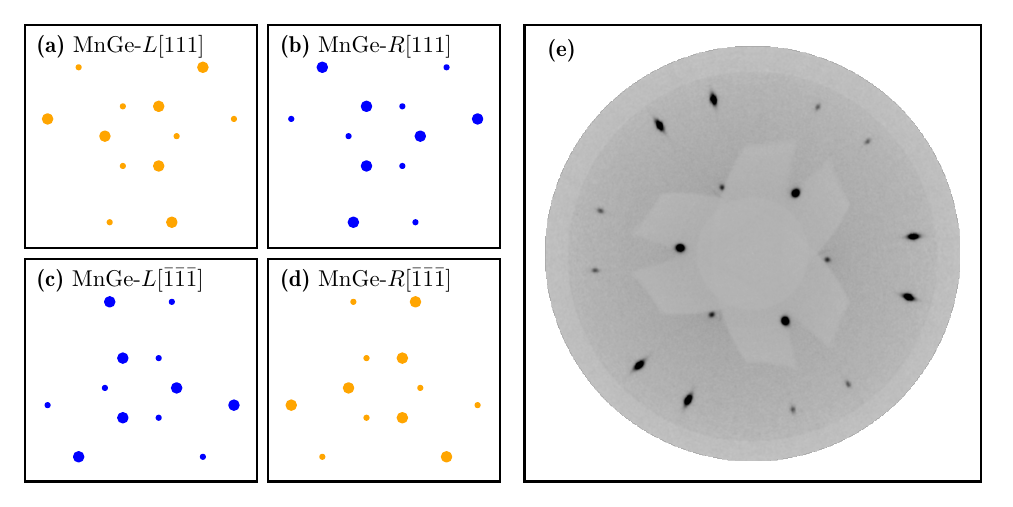}
	\caption{(a)--(d) Simulated [111]-centered pole figures of MnGe$\{210\}$ for MnGe-$L[111]$,  MnGe-$R[111]$, MnGe-$L[\bar{1}\bar{1}\bar{1}]$, and MnGe-$R[\bar{1}\bar{1}\bar{1}]$, respectively. The different spot size represents the intensity pairs given in Table~\ref{tab:210_intensities}. (e) Observed pole figure extracted for radius $q=\SI{29.2}{\per\nano\metre}$ within a width of \SI{0.7}{\per\nano\metre}. Note that the intensity distribution is only explained by superposition of (a) and (d) or (b) and (c).}
	\label{fig:210polefigs}
\end{figure*}

The $P2_{1}3\rightarrow R3$ reduction in symmetry due to the rhombohedral distortion causes an additional shift in both the position and intensity of the peaks. The shifts in peak location are captured by our refinement of the XRD-RSMs, but the intensity shift that leads to four different $\{210\}$ intensities is a small correction. To explain Fig.~7(e) of the main text, we only need to consider the intensities presented in Table~\ref{tab:210_intensities}. We show calculated pole figures for each of the four morphologies in Fig.~\ref{fig:210polefigs} along with the data reproduced from Fig.~7(e) of the main text. The pole figure indicates that twins related by inversion are present in the film: a superposition of  either the \mbox{MnGe-$R[111]$} and \mbox{MnGe-$L[\bar{1}\bar{1}\bar{1}]$} or \mbox{MnGe-$L[111]$} and \mbox{MnGe-$R[\bar{1}\bar{1}\bar{1}]$} simulated pole figures reproduces the angular dependence of the observed spot intensities in the [111]-centered pole figure in Fig.~\ref{fig:210polefigs}(e). The presence of in-plane rotational twinning is ruled out, since such a pair does not reproduce the correct diffraction intensities. Furthermore, the domains must appear in chiral pairs ($L$ and $R$), given the observation of left- and right handed helimagnetic states in SANS~\cite{Kanazawa_2017} which is consistent with our structural measurements.

\begin{table}[H]
\centering
\caption{Cubic B20 MnGe XRD peak intensities for the $\{210\}$ family of reflections.}
\label{tab:210_intensities}
\begin{ruledtabular}
\begin{tabular}{lcccccr}
& $(h k \ell)$ & MnGe-$L[111]$ & MnGe-$L[\bar{1}\bar{1}\bar{1}]$ & MnGe-$R[111]$ & MnGe-$R[\bar{1}\bar{1}\bar{1}]$ & \\
\hline
& ($\bar{2}$$\bar{1}$0) & \textcolor{gray}{0.295} & 100.0 & 100.0 & \textcolor{gray}{0.295} & \\
& ($\bar{2}$0$\bar{1}$) & 100.0 & \textcolor{gray}{0.295} & \textcolor{gray}{0.295} & 100.0 & \\
& ($\bar{2}$01) & 100.0 & \textcolor{gray}{0.295} & \textcolor{gray}{0.295} & 100.0 & \\
& ($\bar{2}$10) & \textcolor{gray}{0.295} & 100.0 & 100.0 & \textcolor{gray}{0.295} & \\
& ($\bar{1}$$\bar{2}$0) & 100.0 & \textcolor{gray}{0.295} & \textcolor{gray}{0.295} & 100.0 & \\
& ($\bar{1}$0$\bar{2}$) & \textcolor{gray}{0.295} & 100.0 & 100.0 & \textcolor{gray}{0.295} & \\
& ($\bar{1}$02) & \textcolor{gray}{0.295} & 100.0 & 100.0 & \textcolor{gray}{0.295} & \\
& ($\bar{1}$20) & 100.0 & \textcolor{gray}{0.295} & \textcolor{gray}{0.295} & 100.0 & \\
& (0$\bar{2}$$\bar{1}$) & \textcolor{gray}{0.295} & 100.0 & 100.0 & \textcolor{gray}{0.295} & \\
& (0$\bar{2}$1) & \textcolor{gray}{0.295} & 100.0 & 100.0 & \textcolor{gray}{0.295} & \\
& (0$\bar{1}$$\bar{2}$) & 100.0 & \textcolor{gray}{0.295} & \textcolor{gray}{0.295} & 100.0 & \\
& (0$\bar{1}$2) & 100.0 & \textcolor{gray}{0.295} & \textcolor{gray}{0.295} & 100.0 & \\
& (01$\bar{2}$) & 100.0 & \textcolor{gray}{0.295} & \textcolor{gray}{0.295} & 100.0 & \\
& (012) & 100.0 & \textcolor{gray}{0.295} & \textcolor{gray}{0.295} & 100.0 & \\
& (02$\bar{1}$) & \textcolor{gray}{0.295} & 100.0 & 100.0 & \textcolor{gray}{0.295} & \\
& (021) & \textcolor{gray}{0.295} & 100.0 & 100.0 & \textcolor{gray}{0.295} & \\
& (1$\bar{2}$0) & 100.0 & \textcolor{gray}{0.295} & \textcolor{gray}{0.295} & 100.0 & \\
& (10$\bar{2}$) & \textcolor{gray}{0.295} & 100.0 & 100.0 & \textcolor{gray}{0.295} & \\
& (102) & \textcolor{gray}{0.295} & 100.0 & 100.0 & \textcolor{gray}{0.295} & \\
& (120) & 100.0 & \textcolor{gray}{0.295} & \textcolor{gray}{0.295} & 100.0 & \\
& (2$\bar{1}$0) & \textcolor{gray}{0.295} & 100.0 & 100.0 & \textcolor{gray}{0.295} & \\
& (20$\bar{1}$) & 100.0 & \textcolor{gray}{0.295} & \textcolor{gray}{0.295} & 100.0 & \\
& (201) & 100.0 & \textcolor{gray}{0.295} & \textcolor{gray}{0.295} & 100.0 & \\
& (210) & \textcolor{gray}{0.295} & 100.0 & 100.0 & \textcolor{gray}{0.295} & \\
\end{tabular}
\end{ruledtabular}
\end{table}

\clearpage

\section*{Structural and magnetic characterization of the 6-nm CrSi sample}
\begin{figure*}[!htb]
	\centering
	\includegraphics[scale=1]{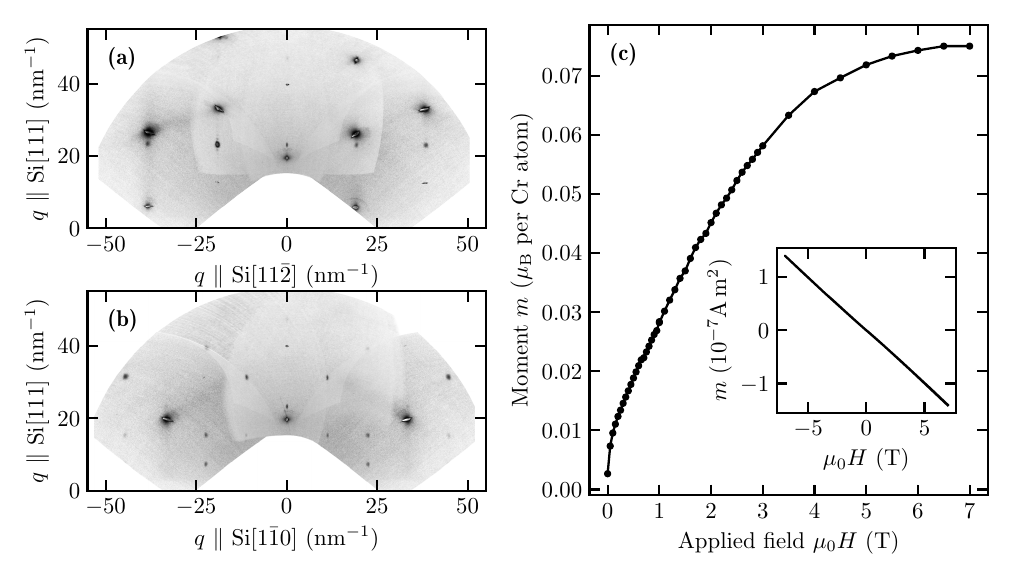}
	\caption{Slices through the XRD-RSM along the (a) Si$[11\bar{2}]$ and (b) Si$[1\bar{1}0]$ in-plane directions of the \SIdash{6}{\nano\metre} CrSi film. (c) Upper estimate of the film magnetization at \SI{5}{\kelvin}. The inset shows the total moment of the film and Si substrate.}
	\label{fig:CrSi}
\end{figure*}

In Figs.~\ref{fig:CrSi}(a)--\ref{fig:CrSi}(b), we show XRD-RSM slices for the \SIdash{6}{\nano\metre} CrSi sample. In the same way as the MnGe samples, this film was refined to the rhombohedrally distorted B20 structure. The lattice parameter was $a = \SI{0.4658}{\nano\metre}$ and the rhombohedral angle was $\alpha = \ang{90.60}$. We also show a preliminary magnetization curve for this sample in Fig.~\ref{fig:CrSi}(c) with field applied parallel to CrSi$[10\bar{1}]$ at \SI{5}{\kelvin}. The inset shows that the signal from the entire sample is largely dominated by the diamagnetism of the Si substrate. To attempt to estimate the film magnetization, we subtracted a linear susceptibility fit to the data above \SI{6}{\tesla}. We therefore obtain an upper bound for the CrSi moment of approximately $0.075\mu_{\mathrm{B}}$ per Cr atom at \SI{7}{\tesla} and \SI{5}{\kelvin}, which is in agreement with the small moments reported elsewhere~\cite{Radovskii_1965, Mishra_2019, Kousaka_2014, Banik_2020}. It is also possible that some of the moment that we attribute to CrSi is coming from paramagnetic dopant atoms in the substrate.

\clearpage
\section*{Transport data of MnGe thin films}
\begin{figure*}[!htb]
	\centering
	\includegraphics[scale=1]{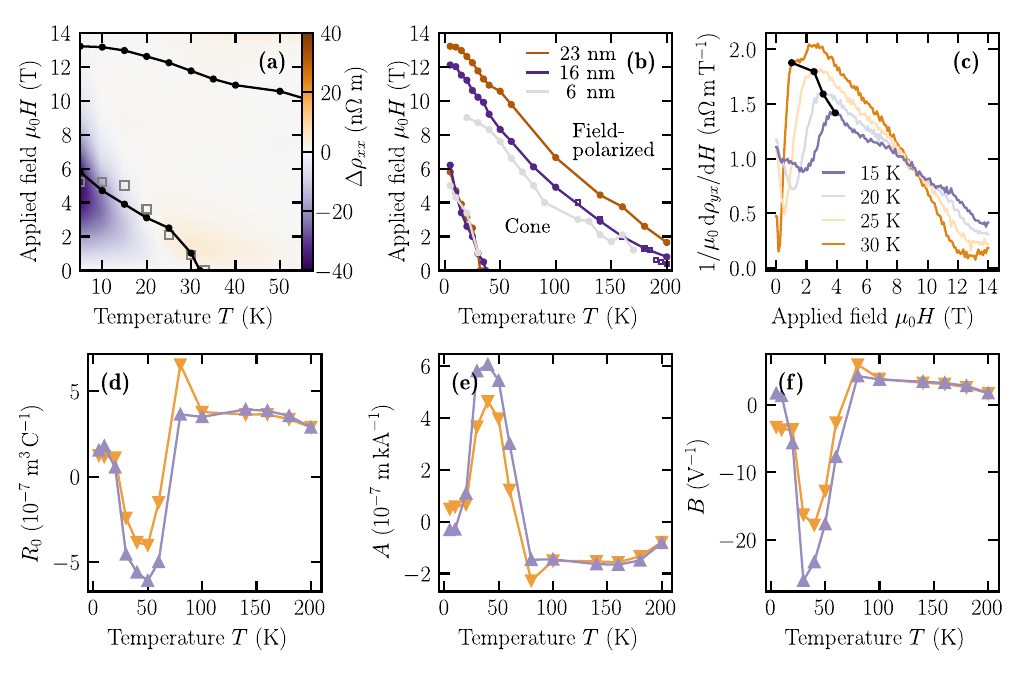}
	\caption{(a) Phase diagram for the \SIdash{22.9}{\nano\metre} sample with colormap showing $\Delta \rho_{xx}$. (b) Full phase diagram shown for \SIdash{22.9}{\nano\metre}, \SIdash{16.4}{\nano\metre} and \SIdash{6.2}{\nano\metre} thick MnGe films. (c) Illustration of the $H_{\mathrm{c}_1}$ transition as determined by $\dv*{\rho_{yx}}{H}$. (d)--(f) Fitting parameters resulting from the application of Eq.2 from the main text to the transverse resistivity. The up (down) triangle markers denote increasing (decreasing) field branches.}
	\label{fig:resistivity}
\end{figure*}

In Fig.~\ref{fig:resistivity}(a), we show the phase diagram for the \SIdash{22.9}{\nano\metre} sample with the colormap depicting $\Delta \rho_{xx}$, to complement Fig.~11(b) of the main text. The phase diagram for \SIdash{22.9}{\nano\metre}, \SIdash{16.4}{\nano\metre} and \SIdash{6.2}{\nano\metre} MnGe samples is given in Fig.~\ref{fig:resistivity}(b) for the full temperature range. While the values of $H_{\mathrm{c}_1}$ are similar, $H_{\mathrm{c}_2}$ is reduced in the thinner films. The transition into the cone phase from the disputed triple-$Q$ or multidomain state is determined from a maximum in $\dv*{\rho_{yx}}{H}$ as demonstrated in Fig.~\ref{fig:resistivity}(c). This feature disappears above \SI{32}{\kelvin} [cf. Fig.~10(c) of the main text]. The fitting parameters $R_0$, $A$, and $B$ of Eq.~2 of the main text used to extract $\rho_{yx}^{\rm{other}}$ from the transverse resistivity are given in Fig.~\ref{fig:resistivity}(d)--\ref{fig:resistivity}(f).

